\documentclass[11pt]{article}

\usepackage[a4paper,margin=2cm,top=2.5cm,bottom=3cm]{geometry}
\usepackage[T1]{fontenc}
\usepackage[utf8]{inputenc}
\usepackage{lmodern}
\usepackage{microtype}
\usepackage{amsmath,amssymb,amsfonts}
\usepackage{booktabs}
\usepackage{graphicx}
\graphicspath{{figures/}}
\usepackage[font=small,labelfont=bf]{caption}
\usepackage{xcolor}
\usepackage{float}
\usepackage{placeins}
\usepackage[title]{appendix}

\usepackage{authblk}

\usepackage[numbers,sort&compress]{natbib}
\usepackage[colorlinks=true,linkcolor=blue!60!black,citecolor=blue!60!black,urlcolor=blue!60!black]{hyperref}

\begin{document}

\title{AI evaluators favor job applications written by certain AI models, independent of candidate quality}

\author[1,*]{Nicol\`{o} Pagan}
\author[1]{Michael Andreas Ebling}
\author[2]{Stefania Ionescu}
\author[1]{Anik\'{o} Hann\'{a}k}

\affil[1]{Universit\"{a}t Z\"{u}rich, Zurich, Switzerland}
\affil[2]{ETH Z\"{u}rich, Zurich, Switzerland}
\affil[*]{Corresponding author: \href{mailto:nicolo.pagan@uzh.ch}{nicolo.pagan@uzh.ch}}

\date{\today}

\maketitle

\begin{abstract}
Job applicants increasingly use large language models (LLMs) to write
cover letters, while employers deploy LLMs to screen them. In this
all-AI setting, we ask whether the \emph{choice of model} affects
hiring outcomes independently of candidate qualifications. We introduce
a multi-model auditing framework pairing eleven LLM writer agents with
nine API-hosted evaluator agents across ten job descriptions and fifty
candidates per job. Four findings emerge. First, we identify a
\emph{generative preference hierarchy}: a consistent global ranking of
writer models by evaluator acceptance, in which locally hosted models
are systematically disadvantaged regardless of their license,
compounded by self-preference bias where evaluators favour their own
outputs. Second, evaluator models differ substantially
in how much bias they introduce based on applicants' tool choice,
offering actionable guidance for fairer screening design. Third, this
hierarchy has measurable competitive consequences:
in the most adverse pairing, approximately two (1.8) out of
twenty-five qualified candidates are filtered out purely due to model
mismatch. Fourth,
a common mechanism explains the patterns above: writer preference and
self-preference trace to a combination of how much each writer leans
on specific stylistic cues and how strongly evaluators, often
unanimously, reward or penalise those same cues. A further check finds
little evidence that an unmeasured ``writer fingerprint'' remains once
these same features are accounted for. We discuss implications for
algorithmic fairness in employment, regulation of AI-assisted
screening, and the design of auditing frameworks that can be reapplied
as the model landscape evolves.

\end{abstract}

\section{Introduction}\label{sec1}

Large language models (LLMs) are now the default technology behind
both sides of the hiring pipeline: a majority of job seekers report
using AI tools to prepare application materials
\cite{resumebuilder_2023}, and employers increasingly deploy LLMs to
screen the applications that result \cite{resumebuilder_2024}. Hiring,
even before LLMs, had long been scrutinized for demographic
bias \cite{raghavan_mitigating_2020,bertrand_are_2004,wilson_building_2021},
and this scrutiny has carried over directly: LLM-based retrieval still
favors white- and female-associated names, with Black male names most
disadvantaged \cite{wilson2024gender}; gains on gender and race have
not extended to educational background \cite{iso2025evaluating}; and
LLM-based evaluators introduce new, opaque discrimination vectors,
inconsistent fairness metrics, and a perpetuation of the biases seen in
human recruiters \cite{albaroudi_comprehensive_2024,
anzenberg_evaluating_2025, veldanda_are_2023, calders_building_2009,
hardt_equality_2016}. Demographic characteristics remain the primary
axis of unfairness studied in LLM-mediated hiring so far.

This is not, however, the only source of inequity in LLM-mediated
hiring: because applicants and employers now draw on the same
landscape of LLM products, a further axis emerges, one rooted not in
\emph{who} a candidate is, but in \emph{which tools} the candidate and
their employer happen to use. This is a technological stratification.

This axis has roots in labor market signaling theory
\cite{spence_job_1973}: application materials are costly, credible
signals of candidate quality precisely because only high-ability
workers find them worth producing, and generative AI disrupts this by
collapsing that cost. On the applicant side, this lowers effort costs
but may also homogenize output quality and erode the informational
content of applications: \citet{wiles_generative_2025} find that
providing both sides of the market with generative AI increased
application volume without improving matching efficiency, and
\citet{awasthi_resumegenai_2025} show that LLM-driven feedback forces
structural homogenization of application materials. \citet{galdin_making_2025}
provide direct evidence: analyzing data from Freelancer.com, they show
that widespread AI use turns written applications into ``cheap talk,''
collapsing their informational value and causing high-ability workers
to be hired 19\% less often.

Once both sides of the pipeline are AI-mediated, a further failure mode
emerges: LLMs systematically favour outputs that resemble their own.
\citet{laurito_aiai_2025} show LLMs assign higher scores to
LLM-generated text over human-written text across decision contexts,
which \citet{panickssery_llm_2024} link to a model's capacity for
implicit self-recognition; \citet{saraf2025quantifying} find the effect
can also be driven by label alone, with a model-name tag shifting
preference share by up to 50\% regardless of true origin and in an
evaluator-specific direction (Claude favoured text labelled as its own;
Gemini favoured text falsely attributed to Claude), suggesting
self-preference has a brand-specific component beyond self-recognition
alone. In hiring specifically, \citet{xu_ai_2026} document
this same self-preference directly: LLM evaluators favour resumes they
generated themselves over human-written equivalents 67--82\% of the
time, translating into a 23--60\% shortlisting premium for candidates
who happen to use the same model as the evaluator, and establishing
AI--AI bias as a concrete labor market risk. However, their one test of a
genuinely different writer model is limited and heterogeneous,
with some evaluators showing no consistent preference at all.

Across this literature, demographic bias has drawn by far the most
scrutiny and technological stratification the least: where AI-AI
dynamics are studied at all, the question asked is usually whether
LLMs prefer AI-generated text over human text in general, or, at most,
an evaluator's preference for its own resumes against one alternative
writer, a narrow case of self-preference that says little about how
genuinely different models compare. What remains unknown is
\emph{which} LLMs are preferred by \emph{which} evaluators across the
\emph{full} space of models in use, and what competitive consequences
follow when all applicants rely on AI. This paper addresses that
question, extending prior sparse and narrow comparisons to a full pairwise audit
across the space of models in active use. To do so, we introduce a
\emph{multi-model
auditing framework} that pairs eleven LLM writer agents with nine
API-hosted evaluator agents across ten job descriptions and fifty
candidates per job, operating as the three-phase pipeline (generation,
evaluation, analysis), designed
to be reapplied as the model landscape evolves. Four findings
emerge from this framework. \textbf{First}, we
identify a \emph{generative preference hierarchy}: a consistent global
ranking of writer models by evaluator acceptance in which locally hosted
models face a structural disadvantage regardless of their license, and
self-preference bias amplifies these effects at the top of the hierarchy.
\textbf{Second}, evaluator models differ substantially in
how much bias they introduce based on applicants' tool choice, and our
analysis identifies which screening models are least harmful, offering
actionable guidance for fairer deployment. \textbf{Third}, this
hierarchy has measurable competitive consequences:
in a rank-displacement analysis where strong (high-fit) candidates and
weak (moderate-fit) candidates are assigned cover letters from
\emph{different} writer models, that writer-model mismatch alone pushes
weaker candidates ahead of stronger ones at rates up to $+7.3\%$ above
the baseline rate observed when both groups use the \emph{same} writer
model, corresponding to approximately two high-fit candidates per
twenty-five displaced by writer choice alone. \textbf{Fourth}, a common
mechanism explains the patterns above in terms of measurable, low-level
writing properties: feature-level
regression traces generative preference and
self-preference to how much each writer leans on stylistic cues
(semantic fit to the job advertisement, readability), crossed with how
unanimously evaluators reward or penalise them. 

\begin{figure}[!htbp]
  \centering
  \includegraphics[width=\textwidth]{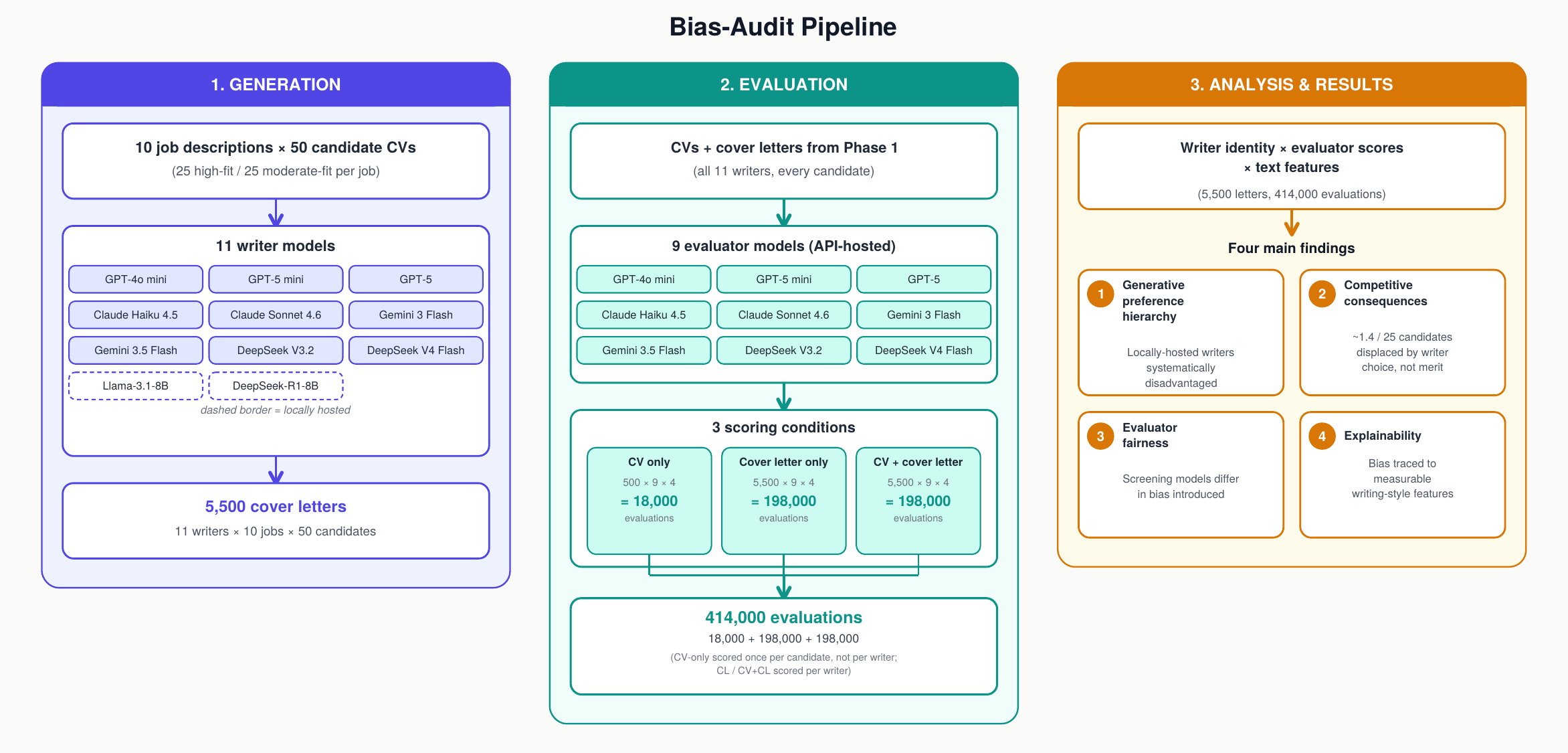}
  \caption{Overview of the three-phase auditing pipeline. In the
    generation phase, eleven writer models (nine API-hosted, two
    locally hosted, dashed border) produce 5,500 cover letters across
    ten job descriptions and fifty candidates per job. In the
    evaluation phase, the nine API-hosted models also serve as
    evaluators, scoring candidates under three conditions (CV only,
    cover letter only, CV + cover letter). CV-only scoring does not
    depend on writer identity, so it is run once per candidate
    (500 candidates $\times$ 9 evaluators $\times$
    4 runs $= 18{,}000$); the other two conditions are run once
    per writer instead (5,500 $\times$ 9 $\times$ 4 $= 198{,}000$ each), for
    $18{,}000 + 198{,}000 + 198{,}000 = 414{,}000$ evaluations in
    total. The analysis phase cross-references writer identity,
    evaluator scores, and text features to produce the four main
    findings previewed on the right.}
  \label{fig:pipeline}
\end{figure}

\section{Results}\label{sec4}

To measure whether writer-model choice shapes hiring outcomes
independently of candidate quality, we build a three-phase auditing
pipeline -- generation, evaluation, and analysis -- summarised in
Figure~\ref{fig:pipeline}. We draw on two publicly available, real-world
datasets, one of CVs and one of job descriptions \cite{resume_kaggle,
jobDesc_kaggle}, and select ten
job postings spanning a wide range of industries and seniority levels
(HR Manager, Primary School Teacher, Accountant, Web Developer, two
distinct UX Designer postings, Chief Financial Officer, Sales Director,
Sales \& Pre-Sales Manager, and Construction Planner; full selection
criteria in Methods, Section~\ref{sec3}). For each job, fifty candidate
CVs are ranked by cosine similarity to the job description and split
into two tiers of twenty-five: \emph{high-fit} and \emph{moderate-fit},
drawn from a middle tier of that ranking (full rationale for this
choice in Methods, Section~\ref{sec3}); the preliminaries below confirm
this split tracks evaluator-perceived quality.

In the \textbf{generation phase}, eleven writer models -- nine
API-hosted models spanning the commercial tools most applicants would
reach for, and two locally hosted models (dashed border in
Figure~\ref{fig:pipeline}) representing applicants without a paid
subscription, who can instead run a small model on their own laptop at
no cost -- each produce one tailored cover letter per candidate,
yielding 5,500 letters in total. In the \textbf{evaluation phase}, the
same nine API-hosted models also serve as evaluators: nine of the
eleven writers therefore double as evaluators of their own and each
other's writing. The two locally hosted models are only ever writers,
never evaluators, reflecting the same asymmetry on the employer side:
an organisation without an established, custom screening pipeline is
far more likely to call a hosted API than to run a locally hosted model
for something as consequential as hiring. Each evaluator scores every candidate under
three conditions: \emph{CV only}, a merit baseline free of any
generative influence; \emph{cover letter only}, which isolates
generative preference from factual grounding; and \emph{CV~+~cover
letter}, the realistic screening scenario. CV-only scoring does not
depend on writer identity, so it is run once per candidate (500
candidates~$\times$~9 evaluators~$\times$~4 runs $=$~18,000
evaluations); the other two conditions are run once per writer instead
(5,500~$\times$~9~$\times$~4~$=$~198,000 each), for
$18{,}000 + 198{,}000 + 198{,}000 = 414{,}000$ evaluations in total. The
\textbf{analysis phase} then cross-references writer identity,
evaluator scores, and measured text features to produce the results
below. Full pipeline detail, including the specific models used and the
generation/evaluation prompts, is in Methods (Section~\ref{sec3}).

\paragraph{What follows.} We proceed in three movements.
Section~\ref{sec:prelim} first confirms the pipeline behaves as
intended: writer letters are comparable and genuinely tailored, and
evaluator scores are reliable and well-calibrated. We then establish
what is happening on both sides of the pipeline:
Section~\ref{sec:preference_hierarchy} identifies a consistent
\emph{generative preference hierarchy} among writers, including
self-preference bias, and Section~\ref{sec:evaluator_fairness} turns to
evaluators, ranking screening models by how much rank-based bias each
one introduces. Section~\ref{sec:competitiveness} then quantifies the
hiring-relevant consequences of that hierarchy through head-to-head win
rates and a candidate rank-displacement analysis. Finally,
Section~\ref{sec:evaluator_explainability} closes by
explaining \emph{why}: both the preference hierarchy and evaluator
neutrality trace to a small set of measurable, low-level writing
features, via feature-level regression.

\subsection{Validating the Experimental Design}\label{sec:prelim}

\paragraph{Writer letters are comparable, yet genuinely tailored.} We
measure each cover letter along several writing-style dimensions --
structure, vocabulary, readability, sentiment and affect, emotion, and
semantic fit to the job ad and CV (full definitions in Methods,
Section~\ref{sec:methods_features}). Across these, the eleven writer
models write in a broadly similar, formal register: Flesch Reading
Ease, a standard 0--100 readability score where higher values mean
simpler text \citep{flesch_new_1948}, falls in the 6--21 range for
every writer, and sentiment is uniformly, almost maximally positive
(VADER $\approx 0.996$ for every writer; full per-feature breakdown in
Appendix~\ref{secA1}). Letter length is similarly consistent: mean word
count ranges from 288 (DeepSeek-R1-8B) to 370 (Gemini~3.5~Flash) across
writers, close to the 350--400 word target specified in the generation
prompt (Section~\ref{sec3}).
Despite this shared register, letters
respond to the specific job: within-job letter similarity (0.68--0.78)
exceeds cross-job similarity (0.46--0.54) for every writer
(Appendix~\ref{secA2}). They also respond to the specific candidate:
letters for high-fit candidates are written in simpler, more readable
language than for moderate-fit candidates, for almost every writer (up
to Cohen's $d = 0.80$), as though writers compensate for a less obvious
fit with more elaborate language (full detail in
Appendix~\ref{secA3}).

\paragraph{Evaluator scores are reliable and well-calibrated.}
Reproducibility across the four independent runs of each (job,
candidate, writer, evaluator) combination is high (mean $\sigma \le
0.50$ points on the 1--10 scale for every evaluator and condition;
Appendix~\ref{secA4}). This justifies the notation used
throughout the rest of the paper: $s_{e,j,c,w}$ denotes the score
evaluator $e$ assigns to candidate $c$'s materials for job $j$ when the
cover letter, if shown, was written by model $w$, taken as the mean
over the four runs; $s_{e,j,c}$ denotes the CV-only score, which does
not depend on a writer. Beyond reproducibility, evaluators are also
well-calibrated: every evaluator recognises the intended
high-fit/moderate-fit gap despite using different absolute scales
($\Delta = 2.25$ on average in the CV-only condition), evaluators agree
strongly with each other on candidate quality (Spearman rank
correlation $\bar\rho = 0.91$--$0.92$ across conditions, on a $-1$ to
$+1$ scale where higher means closer agreement; full definition in
Methods, Section~\ref{sec:methods_calibration}), and each evaluator's own CV-only
ranking agrees with the fixed cosine-similarity split on 86--94\% of
candidates (Appendix~\ref{secA5}).

\subsection{Generative Preference Hierarchy}\label{sec:preference_hierarchy}

Despite writers being broadly similar in style, evaluators do not treat
them as interchangeable. For the same candidate and job, eleven
different cover letters are produced by eleven different writer models,
scored by the same evaluator; we ask whether evaluators are indifferent
about which one they see.

\begin{figure}[!htbp]
  \centering
  \includegraphics[width=0.95\textwidth]{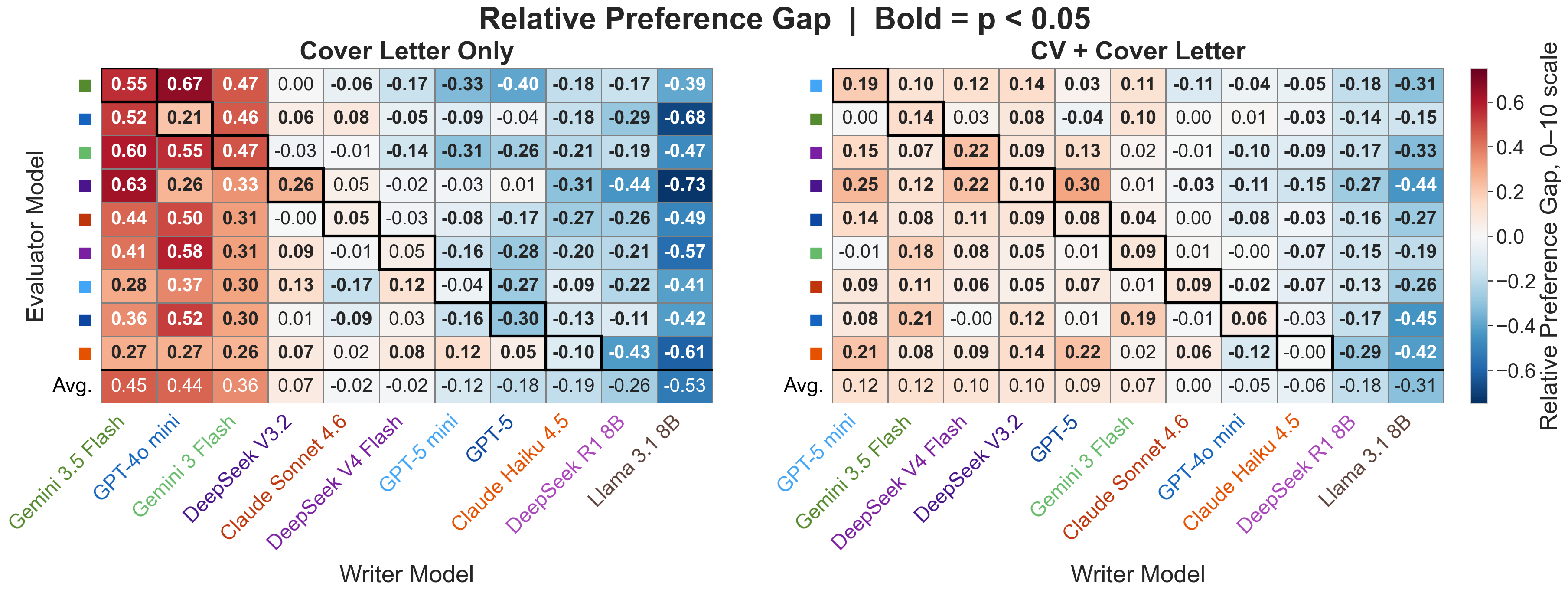}
  \caption{Relative Preference Gap $\Delta_{\mathrm{pref}}(e,w)$: mean
    residual $\tilde{s}_{e,j,c,w}$ averaged over all (job, candidate)
    pairs, for the cover-letter-only (left) and combined CV + Cover letter (right)
    conditions. Columns sorted by decreasing gap; \textbf{bold} values
    indicate $p < 0.05$ (one-sample $t$-test of the 50 per-candidate
    gaps against 0, per cell); true for $80$--$84\%$ of cells, so bold
    is mostly a significance flag: it marks gaps distinguishable from
    zero, regardless of magnitude. Bold
    black-bordered cells (a separate marker) mark
    the self-preference diagonal. Locally hosted models
    are consistently disfavoured, regardless of licensing; self-preference is strongest in the
    combined setting.}
  \label{fig:heatmap}
\end{figure}

\paragraph{Relative Preference Gap.}
To quantify which writer models are systematically favoured or
penalised, we compute for each score the deviation from the evaluator's
mean over all writers for the same candidate: $\tilde{s}_{e,j,c,w} =
s_{e,j,c,w} - \bar{s}_{e,j,c}$, where $\bar{s}_{e,j,c} =
\frac{1}{|W|}\sum_{w} s_{e,j,c,w}$. This per-candidate \emph{preference
residual} removes both evaluator strictness and candidate quality
simultaneously, retaining only how much evaluator $e$ scored writer $w$
above or below its own average for that candidate. Averaging the
residual over all candidates and jobs for a given evaluator--writer
pair $(e, w)$ gives the \emph{Relative Preference Gap}:
\begin{equation}
  \Delta_{\mathrm{pref}}(e, w)
  = \frac{1}{|J||C|}\sum_{j,c} \tilde{s}_{e,j,c,w}.
\end{equation}
A positive value means evaluator $e$ systematically over-rates writer
$w$ relative to its own average; a negative value means it under-rates.

Figure~\ref{fig:heatmap} shows the full matrix for both conditions, and
the result is a clear and consistent \emph{generative preference
hierarchy}. In the cover-letter-only
condition, Gemini~3.5~Flash is the most favoured writer (average gap
$+0.45$), closely followed by GPT-4o-mini ($+0.44$) and Gemini~3~Flash
($+0.36$). Interestingly, GPT-4o-mini outperforms its successor
GPT-5-mini in this condition ($+0.44$ vs.\ $-0.12$), suggesting that
newer architectural iterations do not automatically guarantee superior
stylistic signaling. Locally hosted models are systematically
disfavoured by every evaluator, with Llama-3.1-8B receiving the lowest
average gap of any writer ($-0.53$). The ordering reshuffles
substantially in the combined condition: GPT-5-mini rises to first
($+0.12$), essentially tied with Gemini~3.5~Flash ($+0.12$) and
followed closely by DeepSeek~V4~Flash ($+0.10$), while GPT-4o-mini's
advantage reverses entirely, from $+0.44$ to $-0.05$. Locally hosted
models remain at the bottom across both conditions. As in the
cover-letter-only condition, the magnitude of the preference gaps
shrinks substantially once the CV is also available (the strongest gap
drops from $+0.45$ to $+0.12$), indicating that the generative
preference signal weakens when the CV is also available.

This hierarchy reflects genuine cross-evaluator consensus: evaluators agree well above chance on which
writer wrote the best letter for a given candidate (mean pairwise rank
agreement $\bar\rho = 0.63$ cover-letter-only, dropping to $0.27$ once
the CV is added). That drop does not reflect a loss of reliability,
since evaluators agree on candidate quality just as strongly either way
($\bar\rho = 0.91$--$0.92$); rather, once present the CV supplies most
of the discriminating signal itself, leaving writer identity a smaller,
noisier residual, which is also why the preference gaps above shrink in
the combined condition (full derivation in Appendix~\ref{secA6}).

As global writer preferences attenuate in the combined condition,
self-preference becomes the more prominent signal. The strongest
self-preference in this condition belongs to DeepSeek~V4~Flash
($+0.22$), followed by GPT-5-mini ($+0.19$) and Gemini~3.5~Flash
($+0.14$), while Claude~Haiku~4.5 shows essentially none ($-0.001$).
Section~\ref{sec:evaluator_explainability} traces both the hierarchy
and this self-preference signal back to specific, measurable writing
features.

\subsection{Evaluator Writer-Sensitivity}\label{sec:evaluator_fairness}

\emph{\textbf{Which evaluators are least sensitive to the applicant's
choice of writer model?}} Section~\ref{sec:preference_hierarchy}
established that evaluators, pooled together, show a consistent
generative preference hierarchy. Individually, though, evaluators need
not contribute to that hierarchy equally: Figure~\ref{fig:heatmap}
already breaks that hierarchy down row by row, one row per evaluator,
and this section asks how much each evaluator's own row spreads away
from a neutral midpoint across writers, the practical question being
which evaluator model introduces the least such bias as a function
of applicants' tool choice. An ideal evaluator
would rank writer models randomly around that midpoint,
introducing no systematic preference for any particular model. We
operationalize departure from this ideal as follows.

\begin{figure}[!htbp]
  \centering
  \includegraphics[width=0.95\textwidth]{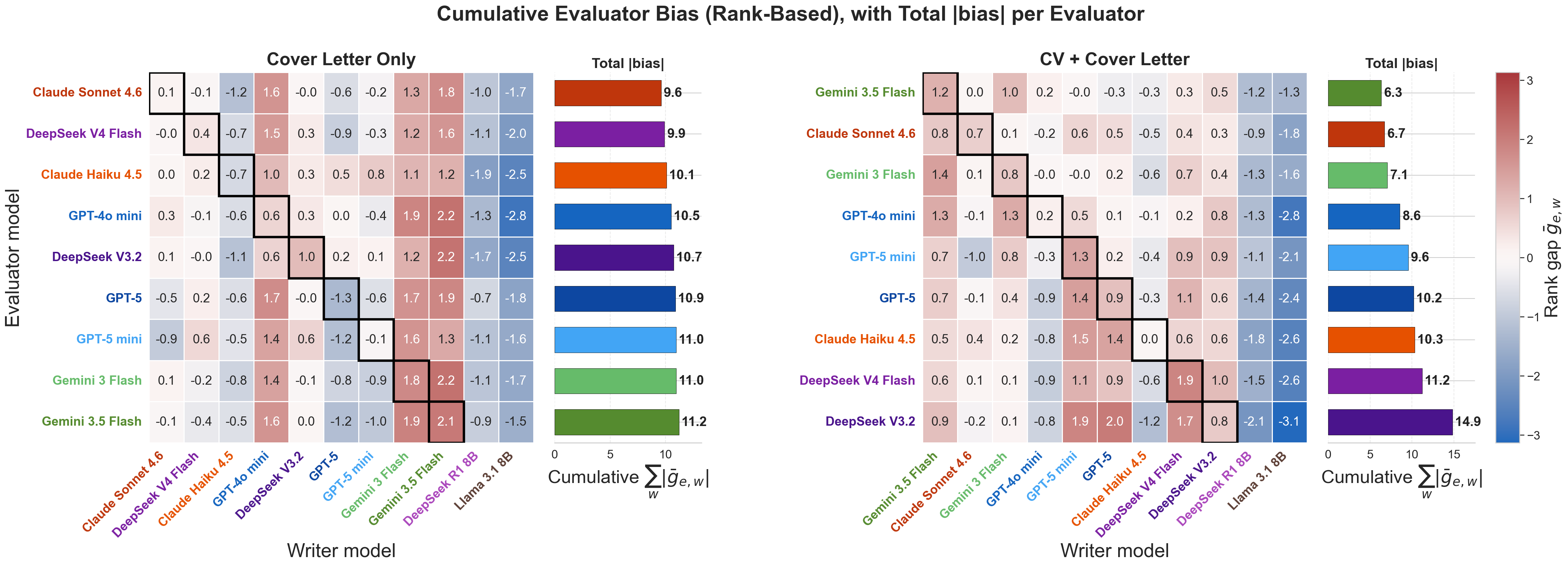}
  \caption{Rank gap $\bar{g}_{e,w}$ as a heatmap (evaluators $\times$
    writers) for the cover-letter-only (left) and combined CV + Cover
    letter (right) conditions; red indicates a writer is over-ranked by
    that evaluator, blue indicates under-ranked. Rows are ordered by
    increasing cumulative $\sum_w|\bar{g}_{e,w}|$ (least to most biased),
    shown as the adjacent bar chart; columns follow the same order for
    every writer that is also an evaluator, so the bold-bordered
    self-preference diagonal (as in Figure~\ref{fig:heatmap}) renders as
    an actual diagonal, with the two writer-only, locally hosted models
    appended at the end. Gemini~3.5~Flash is the most biased evaluator
    in the cover-letter-only condition but the most neutral once the CV
    is also available; DeepSeek~V3.2 becomes the most biased evaluator in
    the combined condition. Locally hosted models are persistently
    under-ranked across all evaluators.}
  \label{fig:fairness}
\end{figure}

Raw scores are not directly comparable across evaluators: beyond
differing in overall strictness (Section~\ref{sec:prelim}), evaluators
also differ in how widely they spread scores across candidates, so a
fixed raw-score gap does not mean the same thing for two evaluators
with different scoring ranges. We therefore compare evaluators on a
rank-based footing instead. For each (job $j$, candidate $c$, evaluator
$e$) triple, the eleven
writer models are ranked by their mean score $s_{e,j,c,w}$ (ties, when
two writers receive the exact same score, are broken by assigning both
their average rank), and the
rank gap for writer $w$ is:
\[
  g_{e,j,c,w} = \mu - \mathrm{rank}_{e,j,c}(w), \quad \mu = \frac{W+1}{2} = 6.
\]
A positive gap means evaluator $e$ ranked writer $w$ above the neutral
midpoint for that candidate; a negative gap means it was ranked below.
We average this gap for each (evaluator, writer) pair across all (job,
candidate) pairs, $\bar{g}_{e,w} = \frac{1}{|J||C|}\sum_{j,c}
g_{e,j,c,w}$, and define the cumulative bias of evaluator $e$ as
$\sum_{w} |\bar{g}_{e,w}|$, summed across all $W$ writer models and
invariant to absolute score levels. Figure~\ref{fig:fairness} shows the
full signed matrix $\bar{g}_{e,w}$ as a heatmap, evaluators against
writers, with each evaluator's cumulative $\sum_w|\bar{g}_{e,w}|$ drawn
as an adjacent bar, so that one can read off both \emph{how biased} an
evaluator is and \emph{towards which writer models}.

In the cover-letter-only condition, Gemini~3.5~Flash (total: 11.2) is
the most biased evaluator, while Claude~Sonnet~4.6 (9.6) is the most
neutral. The decomposition reveals that Gemini~3.5~Flash's bias stems
primarily from over-ranking itself ($+2.1$ above midpoint on average)
and its sibling Gemini~3~Flash ($+1.9$), while under-ranking
Llama-3.1-8B ($-1.5$) and GPT-5 ($-1.2$): a combination of
self-preference and cross-family favouritism. In the combined
condition, the evaluator ordering shifts entirely: Gemini~3.5~Flash
becomes the most neutral (6.3) while DeepSeek~V3.2 becomes the most
biased (14.9). Most evaluators' cumulative bias falls once the CV is
present (five of nine drop by at least 1.4 points), but DeepSeek~V3.2's
instead rises sharply ($10.7\to14.9$), with only DeepSeek~V4~Flash
moving the same direction ($9.9\to11.2$, far more modestly). The
per-writer decomposition for DeepSeek~V3.2 shows
GPT-5 ($+2.0$) and GPT-5-mini ($+1.9$) as the most over-ranked
writers, while Llama-3.1-8B ($-3.1$) and DeepSeek-R1-8B ($-2.1$) are
the most persistently under-ranked. This locally hosted disadvantage is
not driven by any single evaluator but forms a structural floor in the
bias distribution, consistent with the generative preference hierarchy
documented in Section~\ref{sec:preference_hierarchy}.

\subsection{Writer Model Competitiveness}\label{sec:competitiveness}

We now examine the competitive consequences of the preference hierarchy
identified in Section~\ref{sec:preference_hierarchy} and the per-evaluator
bias documented in Section~\ref{sec:evaluator_fairness}, focusing on the combined CV + Cover letter condition.
Two analyses are presented: a head-to-head win rate
tournament measuring which writer models consistently outperform others
across evaluators, and a rank displacement analysis quantifying how
often a moderate-fit candidate using a stronger writer model leapfrogs
a high-fit candidate using a weaker one. Both confirm that model choice
has measurable hiring consequences independent of candidate quality.

\begin{figure}[!htbp]
  \centering
  \includegraphics[width=0.95\textwidth]{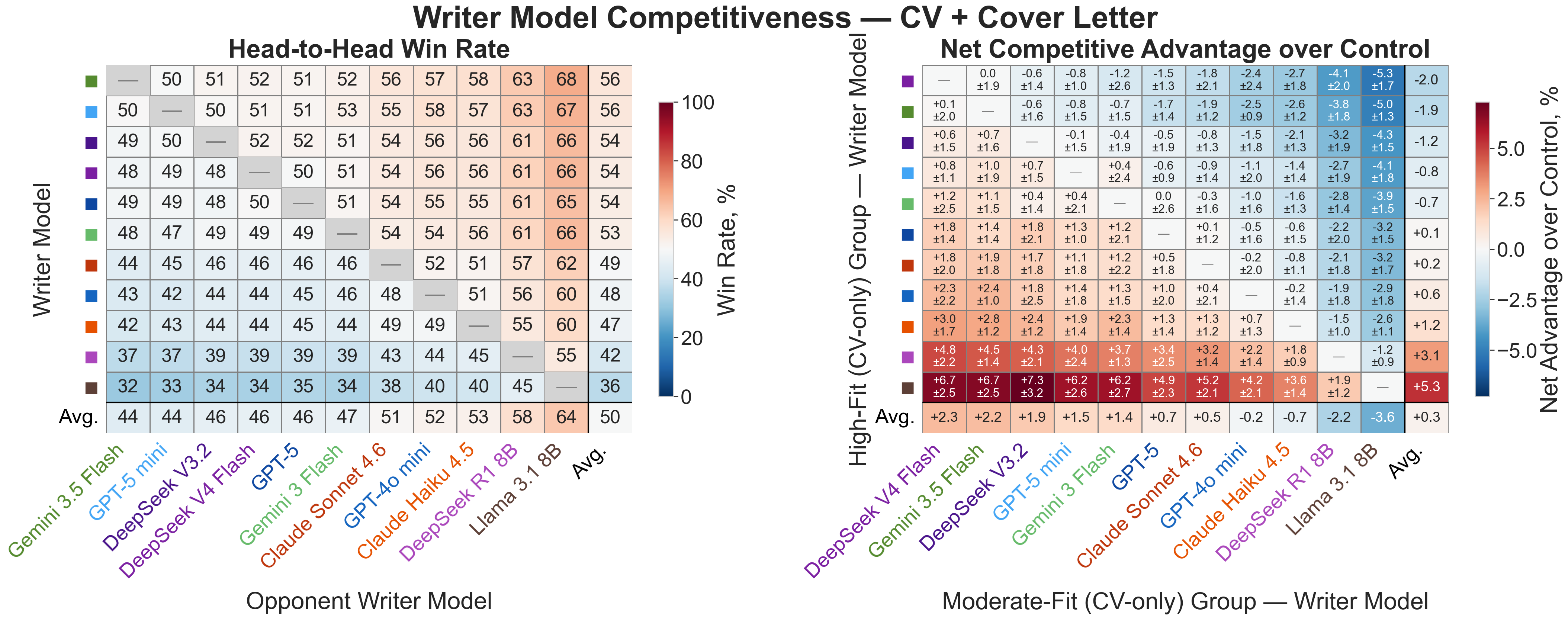}
  \caption{Both panels refer to the combined CV + Cover letter condition.
    Left: head-to-head win rates $\mathrm{WR}(w_1, w_2)$; triples where
    the evaluator matches either writer are excluded to remove
    self-evaluation confounds.
    Gemini~3.5~Flash leads (56\%), closely followed by GPT-5-mini and
    DeepSeek~V3.2 (56\% and 54\%); locally hosted models consistently lose.
    Right: net displacement rate $\Delta R(w_2, w_1)$ (percentage points
    above the same-writer baseline) for the scenario where all high-fit
    candidates (defined as the top 25 by each evaluator's own CV-only
    ranking, recomputed per evaluator, job, and run) use writer
    $w_1$ (rows), and all moderate-fit candidates (bottom 25 by the same
    ranking) use writer $w_2$ (columns). Diagonal cells (grey ``--'') are
    the control condition, $w_1 = w_2$, by definition 0 and omitted.
    Off-diagonal cells show the mean net rate and, beneath it, $\pm$ its
    standard deviation across the nine evaluators.
    In the most adverse pairing (Llama-3.1-8B vs.\ DeepSeek~V3.2),
    approximately two (1.8) out of twenty-five high-fit candidates are displaced
    solely due to model mismatch.}
  \label{fig:win_nca}
\end{figure}

\paragraph{Head-to-head win rates.}
For every ordered pair of writer models $(w_1, w_2)$, we compute the
fraction of (job $j$, candidate $c$, evaluator $e$) triples in which $w_1$ scores
strictly higher than $w_2$, with ties counting as 0.5. Triples where
the evaluator coincides with either writer are excluded, isolating
cross-model assessments:
\[
  \mathrm{WR}(w_1, w_2) = \frac{\mathrm{wins}(w_1, w_2) +
    0.5 \cdot \mathrm{ties}(w_1, w_2)}{\mathrm{matches}(w_1, w_2)} \times 100\%.
\]
The left panel of Figure~\ref{fig:win_nca} shows the full matrix,
sorted by each model's average win rate. Gemini~3.5~Flash leads at 56\%,
closely followed by GPT-5-mini (56\%) and DeepSeek~V3.2 (54\%).
Both locally hosted models occupy the bottom of the ranking
(DeepSeek-R1-8B at 42\%, Llama-3.1-8B at 36\%), with Claude~Haiku~4.5
(47\%) the weakest API-hosted model, confirming a systematic competitive
disadvantage specific to locally hosted models: licensing category
alone does not explain it.

\paragraph{Net competitive advantage.}
Win rates measure pairwise dominance in scoring, but the hiring
consequence that matters is rank displacement: does a better-performing
writer model cause a moderate-fit candidate to leapfrog a high-fit one?
The right panel of Figure~\ref{fig:win_nca} quantifies this directly:
for each combination of evaluator model, job, and evaluation run
independently, all 50 candidates are first ranked by the evaluator's
own CV-only scores. The top 25 by this ranking are designated
\emph{high-fit (CV-only)} and the bottom 25 \emph{moderate-fit
(CV-only)} -- recomputed per evaluator, job, and run, unlike the
cosine-similarity-based split used elsewhere. All high-fit candidates
are then assigned the cover letter of writer $w_1$ (rows), all
moderate-fit candidates that of writer $w_2$ (columns); a displacement
event is a moderate-fit candidate ascending into the high-fit group
once the letter is added, pushing out a high-fit candidate who would
have been shortlisted on CV alone.

Some baseline displacement is expected even without any model
asymmetry, since the two tiers' score distributions overlap
(Figure~\ref{fig:strictness_by_condition}(a)); we isolate the
writer-model effect by subtracting the same-writer baseline rate
$R_{\mathrm{raw}}(w_1, w_1)$ from the raw displacement rate
$R_{\mathrm{raw}}(w_2, w_1)$, giving the net displacement rate
\begin{equation}
  \Delta R(w_2,w_1) = R_{\mathrm{raw}}(w_2, w_1) - R_{\mathrm{raw}}(w_1, w_1).
\end{equation}
Row $w_1$, column $w_2$ thus gives the net rate (percentage points) by
which moderate-fit candidates using $w_2$ displace high-fit candidates
using $w_1$, averaged across the nine evaluators; the standard
deviation shown beneath each cell is evaluator-to-evaluator spread.

In the most adverse pairing, high-fit candidates using Llama-3.1-8B and
moderate-fit candidates using DeepSeek~V3.2 (a top-3 model by win
rate), the net displacement rate reaches $+7.3\% \pm 3.2\%$ --
approximately two of twenty-five high-fit candidates per job reversed
solely by model mismatch, invisible to an employer whose LLM screener
passes only top-ranked candidates to human review. This figure is
already noise-subtracted against a substantial baseline: same-writer
scoring noise alone displaces $7.6$--$9.5\%$ of candidates depending on
the writer (mean $8.2\%$), so the net effect roughly doubles that
floor. The spread also means individual evaluators can go well beyond
the pooled mean: GPT-4o-mini and DeepSeek~V3.2 both reach up to
$+12.8\%$ ($\approx 3.2$ of 25) on specific pairings, detailed in
Appendix~\ref{secB}.

\subsection{Explaining Preference Through Writing-Style
Features}\label{sec:evaluator_explainability}

Section~\ref{sec:preference_hierarchy} showed \emph{that} writer
preference and self-preference exist; Section~\ref{sec:evaluator_fairness}
measured \emph{how much} each evaluator's rankings depart from
neutrality; and Section~\ref{sec:competitiveness} translated both into
concrete hiring stakes. This section asks \emph{why}: feature-level
Ridge regression traces the preference hierarchy and
self-preference to how much writers lean on specific, measurable
writing-style features, crossed with how much evaluators reward or
penalise those same features.

\begin{figure}[!htbp]
  \centering
  \includegraphics[width=0.95\textwidth]{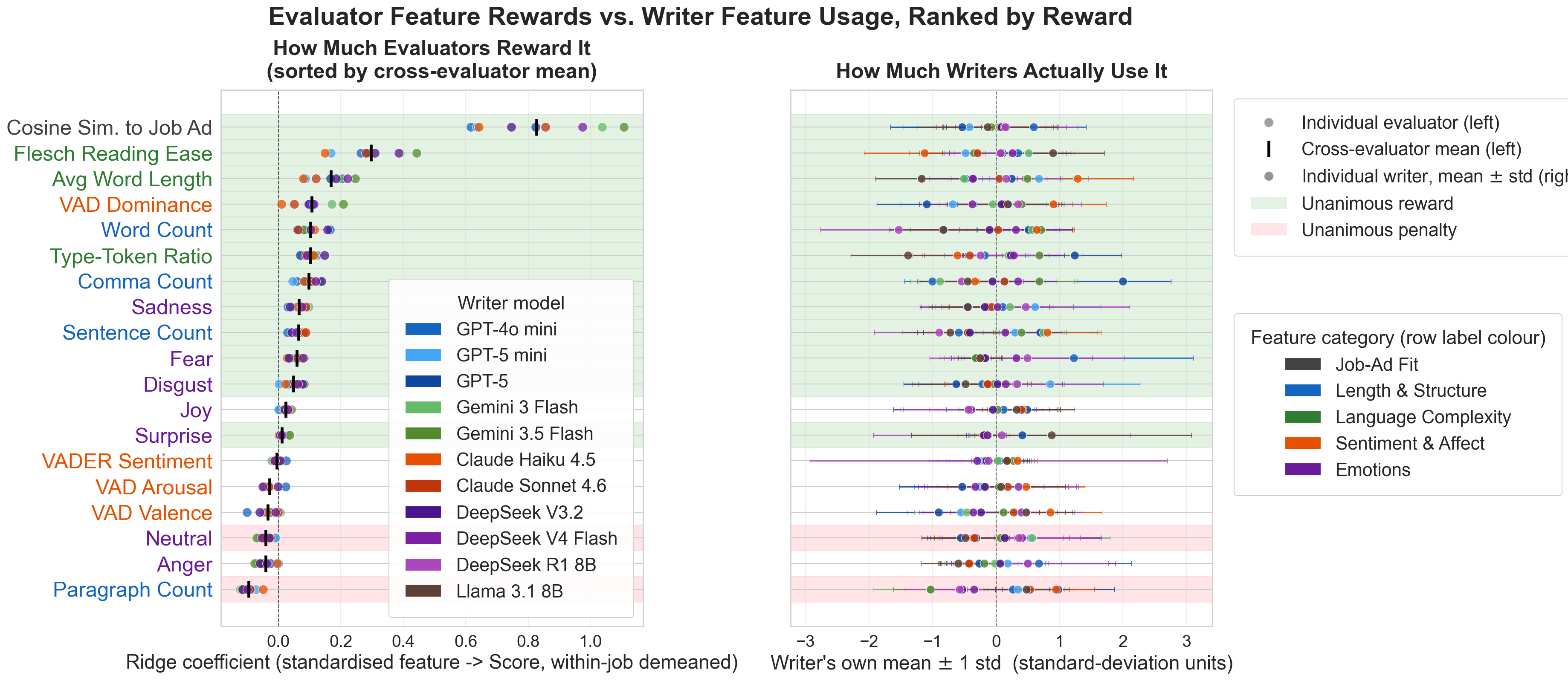}
  \caption{Left: cross-evaluator Ridge coefficient for each feature
    (cover-letter-only condition; Cosine Similarity to CV excluded, see
    text), one dot per evaluator plus a bold bar for the
    cross-evaluator mean, ranked top to bottom by that mean; rows
    shaded when all nine evaluators agree in sign (green = unanimous
    reward, red = unanimous penalty). Right: each writer's own mean use
    of that feature (mean $\pm$ 1 std), standardised to the same units.
    Row labels are coloured by feature category. As throughout,
    Cosine Similarity to Job Ad/CV measures the \emph{generated letter's}
    embedding against the job ad/CV. It is not the job ad and CV
    measured against each
    other (Figure~\ref{fig:writer_features}). Cosine Similarity to Job Ad
    is both the most strongly and the most unanimously rewarded feature
    by a wide margin; Paragraph Count, Anger, and Neutral are the
    clearest unanimous penalties.}
  \label{fig:common_pref_ranked}
\end{figure}

\subsubsection{Feature-Level Regression}\label{sec:explain_regression}

\paragraph{Linking feature-level preference to writer-level outcomes.}
For each evaluator, we fit a Ridge regression of score on the
(standardised) writing-style features -- an L2-regularized linear
regression that shrinks coefficients toward zero to stay well-behaved
under the features' mutual correlation \citep{hoerl_ridge_1970}. The
resulting coefficients (Figure~\ref{fig:common_pref_ranked}) link the
Relative Preference Gap
(Section~\ref{sec:preference_hierarchy}, a writer-level outcome) to its
feature-level cause, decomposing evaluator preference to the feature
level in the cover-letter-only condition -- isolating the letter's
stylistic effect on evaluator preference requires removing the CV,
whose presence would otherwise dominate and confound any purely
stylistic signal -- across the twenty writing-style
features we measure: structure (word, sentence, paragraph, and comma
counts), vocabulary (average word length, type-token ratio), readability
(Flesch Reading Ease), sentiment (VADER), affect (VAD valence, arousal,
dominance), emotion (seven categories), and semantic fit (cosine
similarity to the job ad and to the CV; full definitions in Methods,
Section~\ref{sec:methods_features}). Cosine Similarity to CV is deliberately
excluded from this condition's regression, leaving nineteen: the
evaluator's prompt never contains the CV when scoring the cover letter
alone, so any coefficient the model assigned it would reflect a
spurious correlate of something else
(Appendix~\ref{secC2} revisits this feature once the CV is actually
visible). For each of the nineteen, the left panel shows how strongly
and how unanimously the nine evaluators reward it, ranked top to bottom
by cross-evaluator mean Ridge coefficient; the right panel shows how
much each of the eleven writers actually uses that feature,
standardised to the same scale. The mechanism linking the two panels is
direct: a writer that leans unusually hard into a feature evaluators
reward strongly and unanimously should show a correspondingly strong
Relative Preference Gap, and a writer that under-uses such a feature
should be penalised for it. Cosine Similarity to Job Ad is by a wide margin
the most strongly and unanimously rewarded feature of the nineteen
(cross-evaluator mean $\approx 0.83$, nearly three times the next-highest,
Flesch Reading Ease at $\approx 0.30$), and GPT-4o-mini is, by a wide
margin, the writer that leans furthest into it ($z \approx +0.59$
standard deviations above the writer-pool mean, the single most
extreme writer--feature combination in the figure), exactly the
mechanism behind its high cover-letter-only Relative Preference Gap
($+0.44$, second only to Gemini~3.5~Flash's $+0.45$; Figure~\ref{fig:heatmap}).
Writers that
under-use a strongly rewarded feature are penalised symmetrically:
Claude~Haiku~4.5 has by far the lowest Flesch Reading Ease of any
writer ($z \approx -1.13$), consistent with its comparatively weak
overall standing ($-0.19$); it partially compensates by leaning
hardest into two other unanimously rewarded features, Average Word Length
($z \approx +1.29$, the highest of any writer) and VAD Dominance
($z \approx +0.90$, also the highest), keeping it from falling further.
DeepSeek-R1-8B's low standing in the same figure ($-0.26$) is more one-sided: its word
count is the lowest of any writer by a wide margin ($z \approx -1.54$)
on a feature that is itself unanimously rewarded, with no comparably
strong compensating feature elsewhere in its profile.

A finer-grained, per-evaluator view of the Cosine Similarity to Job Ad row
specifically (Appendix~\ref{secC1}, Figure~\ref{fig:semantic_fit_clonly})
surfaces a further asymmetry: GPT-4o-mini's own Ridge coefficient for
that feature ($+0.60$) is the \emph{lowest} of all nine evaluators,
less than half of Gemini~3.5~Flash's and Gemini~3~Flash's ($+1.10$,
$+1.04$): its dominant standing on this feature traces to how
extreme its own writing is, independent of how generously it personally
rewards the trait. Three within-writer style trade-offs documented in
Appendix~\ref{secA1} -- readability vs.\ average word length, dominance
vs.\ arousal, and Joy vs.\ Neutral tone -- show up in evaluator
preference too, each resolved differently. Writers cannot combine
long, complex words with easy-to-read structure: the two trade off
sharply in how letters are actually written (longer average words
track lower Flesch Reading Ease), yet evaluators reward both
simultaneously regardless, an ideal combination essentially no
writer among the eleven actually produces. Dominance and arousal move
together in how writers write (a more assertive-sounding letter also
reads as more energetic), but that pairing partially reverses in
evaluator preference: every model rewards dominance, yet the ones that
reward it most tend to reward arousal least, favouring ``confident but
composed'' language over ``assertive and energetic.'' Joy and Neutral
tone trade off almost perfectly in how writers write them, close to
mechanical since the two are categories of one probability
distribution; evaluators are under no such constraint, since their
Ridge coefficients on the two are fit independently, yet mirror that
same trade-off even more steeply -- a genuine preference structure, not
an artifact of how the features are measured (full detail and figures
in Appendix~\ref{secC1}).

The same feature-preference
picture extends to the CV~+~Cover~Letter condition with one addition:
Cosine Similarity to CV, excluded from the cover-letter-only analysis,
re-enters and becomes as strongly and unanimously rewarded as Cosine
Similarity to Job Ad ($0.60$ vs.\ $0.60$), together the two most
rewarded features by a wide margin. Full per-feature detail and
figures for this condition are in Appendix~\ref{secC2}.

\paragraph{Self-preference is feature-specific, not predictive of
standing.}
The analysis above explains the general hierarchy; we now turn to
self-preference itself, the signal
Section~\ref{sec:preference_hierarchy} found becomes more prominent
once the CV is added. Here too there is no fixed per-model trait: the
same writer can be the most self-rewarding evaluator on one feature and
among the most self-critical on another, and a writer's own extreme
usage of a feature can dominate its self-preference even when that
writer is not the feature's most generous rewarder (full worked
example in Appendix~\ref{secC2}). Summed across all measured features
into a single number per writer, does this feature-level self-flattery
translate into being more broadly liked? We compare each writer's
\emph{Net Self-Preference} (its own idiosyncratic bias, own taste $-$
average taste, dotted with its own style, summed across all measured
features) against its \emph{Total Preference Received} (that same
writer's actual style dotted with the summed taste of all nine
evaluators, self included): the two are nearly inverted at the
extremes in the cover-letter-only condition, with the model most
critical of its own writing the one the panel likes most overall, and
vice versa (full per-writer numbers in Appendix~\ref{secC2}). So
self-preference does not predict overall standing. The CV changes the
two levels differently: feature-level self-preference is, if anything,
sharper once the CV is present (Appendix~\ref{secC2}), while this
aggregate pattern collapses almost uniformly toward the origin,
consistent with the CV supplying most of the discriminating signal.
Full figure, per-writer numbers for both conditions, and the
relationship to the Relative Preference Gap of
Section~\ref{sec:preference_hierarchy} are in Appendix~\ref{secC2}.

\paragraph{These measurable features are close to sufficient.}
Appendix~\ref{secC3} decomposes evaluators' explainable score variance
into a shared trunk (Job, Substance, Evaluator baseline) and what
remains, and finds little room for an unmeasured ``writer
fingerprint'': measurable Text Features explain almost as much of the
residual as raw Writer identity does ($0.910$ vs.\ $0.911$ in
CV~+~Cover~Letter, a similar pattern in Cover-Letter-Only). The same
appendix also breaks this down per evaluator, connecting it back to
Section~\ref{sec:evaluator_fairness}'s neutrality ranking.

\FloatBarrier
\section{Discussion}\label{sec5}

\paragraph{Who this harms, and why it stays invisible.}
Three things are worth separating about this harm: what the underlying
mechanism actually is, how unevenly it lands even among the people it
affects, and why employers cannot see it happening. Our findings point
to a new axis of unfairness in automated hiring:
technological stratification \cite{gallegos_bias_2024,
albaroudi_comprehensive_2024}. The underlying mechanism is that
evaluator models do not treat writer models equally, regardless of
which candidate stands behind a given letter. That becomes a problem
for people as well as models, because of who ends up using which
writer: applicants without paid API access have only free, locally
hosted models available to them, and that group skews toward
socioeconomic disadvantage even though it does not map cleanly onto
race, gender, or nationality as defined in employment discrimination
law -- unlike demographic bias, this axis targets no legally protected
group. Prior work has shown that AI adoption erodes the informational
content of applications on average \cite{galdin_making_2025}; our
results point to something more specific happening on top of that: the
signal is being actively redirected toward whichever models
better-resourced applicants can afford, because employers' own
screening models reward exactly those models' output.

Even within this group of affected applicants, the harm is not evenly
distributed. Every displacement event (Section~\ref{sec:competitiveness})
cuts two ways at once -- a moderate-fit candidate wrongly admitted, a
high-fit candidate wrongly excluded -- and those two sides do not carry
equal weight. If a moderate-fit candidate gets shortlisted on the
strength of a better cover letter, that is likely to be caught at
interview or review, so the employer's exposure is limited. A high-fit
candidate filtered out because their cover letter was weaker never
gets that far, and for that candidate this is the outcome that
matters: a real opportunity lost, with no recourse and no way of
knowing that model choice decided the outcome, when they might well
believe it came down to their own qualifications.

This asymmetry is compounded by a third problem: invisibility to
employers. Unlike demographic bias, which leaves detectable
statistical signatures in hiring outcome distributions and can in
principle be detected via post-hoc audit, technological stratification
requires prospective auditing of the kind our framework enables.
Employers have no visibility into which tool applicants used and cannot
observe the bias they are inadvertently introducing. Our displacement
estimates are also almost certainly lower bounds: candidates who forgo
AI assistance entirely face a substantially larger
gap \cite{laurito_aiai_2025, xu_ai_2026}, so the within-AI analysis
we report represents the minimum magnitude of the phenomenon.

\paragraph{What employers and regulators can do about it.}
Two things make this tractable: a concrete
regulatory hook already exists, and the mechanism itself is legible,
traceable to measurable features (Section~\ref{sec:evaluator_explainability}).
Regulatory frameworks such as the EU AI
Act \cite{eu_ai_act_2024},
which classify automated hiring tools as high-risk AI systems subject
to conformity assessments, provide a hook for mandating model-of-origin
audits, though current guidance does not yet address this axis
specifically. For employers seeking to act now, our evaluator fairness
analysis offers concrete guidance: models with lower cumulative bias
are less likely to introduce systematic model-origin effects, and
separating the roles of generation support and candidate screening
avoids the power-concentration risk illustrated concretely by
DeepSeek~V3.2's dual role as both a leading self-preferencing writer
and the most biased evaluator (Section~\ref{sec:evaluator_fairness}).
Writing quality and evaluator fairness are not inherently in tension,
though: Gemini~3.5~Flash is a leading writer in both conditions yet
becomes the most neutral evaluator once the CV is present
(Section~\ref{sec:evaluator_fairness}) -- the fairest evaluators in our
audit are not the weakest writers, so employers need not trade off one
for the other when choosing an evaluator model.

Beyond identifying which evaluators are more or less biased, the
feature-level mechanism of
Section~\ref{sec:evaluator_explainability} clarifies what that bias
actually consists of: generative preference traces to measurable,
low-level writing
properties (structure, readability, and semantic fit to the job
advertisement), and a further check (Appendix~\ref{secC3}) finds
little room for an unmeasured writer identity on top of those same
features. Little of the effect looks like some irreducible
reaction to a model's identity as a ``brand.'' Practically, this is
grounds for cautious optimism: because the mechanism is legible,
mitigation does not require reverse-engineering model internals, only
monitoring a tractable set of surface features.

\paragraph{Limitations.}
Four limitations are worth being explicit about: how low absolute
scores in the CV-only condition should be read, an ecological-validity
gap in how CV-job pairings were assigned, what our fixed anonymization
choice does and does not let us say,
and a methodological caveat
specific to the variance-decomposition analysis.

Absolute scores are low in the CV-only condition specifically -- mean
scores around 4 for high-fit candidates and around 2 for moderate-fit
candidates, both below the rubric's own bar for a genuinely qualified
applicant -- because it is the sparsest signal we evaluate, a bare CV
with no persuasive framing; it is not a ceiling on how highly any
evaluator scores these candidates. In the combined condition, high-fit
candidates average $4.83$, and GPT-4o-mini alone rates them $7.10$,
squarely in the rubric's ``strong match'' band. Our central claim does
not depend on the absolute level in any case: it is a within-candidate
comparison, asking whether the same candidate's score moves when only
the writer model changes, and that comparison is well-defined
regardless of how strict or lenient an evaluator is on average. What it
does not establish is that the effect generalises to candidates a human
recruiter would independently call excellent; validating our candidate
pool against human judgment is a natural next step we leave to future
work.

Our CVs and job descriptions are each real, but the pairing between
them is not: no candidate in our pool actually applied for the job
their cover letter was generated against. Real applicants self-select
into jobs they already believe themselves a plausible fit for, which
our cosine-similarity-constructed candidate pool does not replicate,
and a writer model generating a letter for a pairing no real applicant
chose may not fully capture how a genuine, motivated application would
be framed. Our results measure model behaviour under a controlled,
researcher-assigned pairing, not necessarily under the self-selected
pairings a real applicant pool would exhibit.

All CVs are anonymized to remove gender markers, with a fixed
gender-neutral name used throughout \cite{anzenberg_evaluating_2025};
we therefore cannot speak to whether gendered names would change
evaluator behaviour, only to the writer-model effect isolated when they
do not vary.

The variance-decomposition analysis of
Appendix~\ref{secC3} carries two further caveats. A
raw-score variance decomposition and a rank-based fairness metric can
diverge sharply, particularly in noisier evaluation conditions where
scores cluster near decision boundaries, so ranking-level harm to
candidates should not be inferred from a variance audit alone. It also
only controls for each evaluator's own strictness (a fixed intercept),
not for how widely each evaluator spreads scores; since $R^2$ is
computed on pooled residuals, evaluators with wider score spread
implicitly carry more weight in what counts as explained variance, so
those results are best read as representative of the higher-spread
evaluators rather than uniformly true across all nine.

The specific model hierarchy we report will shift as models are updated
or deprecated, and our primary contribution is therefore methodological:
the framework itself is the durable deliverable.
We recommend that organisations conducting impact assessments of AI
hiring tools incorporate periodic replication of this audit as standard
practice, analogous to the periodic demographic bias audits now required
in several jurisdictions. Future work should examine generalisability
across languages and labour markets, whether instruction-tuned
neutral-writing strategies can mitigate generative preference bias, and
whether mandatory disclosure of AI use in applications changes evaluator
behaviour.

\subsection*{Ethical considerations}\label{sec:ethics_discussion}

This study uses only publicly available CV and job description datasets;
no personal data were collected. All candidate profiles were anonymized
prior to analysis. Models were accessed via public APIs under their
respective terms of service.

The findings could in principle inform strategic model selection by
applicants seeking to game screening systems. We consider this risk
modest: model preferences shift as models are updated, and our primary
contribution is a detection framework. It tracks a moving target; it
does not hand applicants a stable gaming recipe. Surfacing generative
preference bias publicly is more likely to
prompt corrective action by employers and regulators than to enable
durable individual gaming. Publicly identifying which evaluator models
are least biased also creates an accountability mechanism that serves
candidates, employers, and regulators alike.

A potential adverse impact is discouraging AI screening adoption,
including tools that may on balance reduce demographic bias relative to
unassisted human screening. Our findings do not imply AI screening is
worse than human screening overall; they identify a specific, currently
invisible and unregulated bias axis that can be mitigated through
thoughtful evaluator selection and periodic auditing.

\section{Conclusion}\label{sec6}

We built a reusable auditing framework that isolates generative
preference bias from candidate quality: eleven writer models (nine
API-hosted, two locally hosted) produce cover letters for fifty
candidates across ten job postings, and nine API-hosted evaluator
models score every candidate under three conditions -- CV only, cover
letter only, and CV plus cover letter -- for 414,000 evaluations in
total (Figure~\ref{fig:pipeline}). Candidate quality is fixed by an
external, evaluator-independent criterion, so every evaluator is judged
against the same ground truth and writer identity is the only thing
that varies within a candidate.

When both applicants and employers use AI in recruitment, the
\emph{which} matters as much as the \emph{whether}. Four findings
follow from this design. First, a persistent generative preference
hierarchy means that equally qualified candidates face different
screening outcomes depending solely on which language model wrote
their cover letter, a bias invisible to the employer and unrecoverable
for the candidate; locally hosted models, run by those without means
to pay for API access, are systematically disadvantaged regardless of
their underlying licence.
Second, evaluator models differ substantially in how much of this bias
they introduce, giving employers an actionable lever even before the
underlying hierarchy itself can be fixed. Third, this hierarchy has
measurable
competitive consequences: writer-model mismatch alone displaces
qualified candidates at a rate comparable to the noise floor itself.
Fourth, how much writers
lean on stylistic cues, crossed with
how evaluators reward or penalise them, accounts for almost
all of the preference hierarchy, and a further check finds little
unaccounted-for ``writer fingerprint'' in evaluator scores: the bias
is legible.
As AI-mediated hiring
scales globally, model-of-origin bias warrants the same regulatory
scrutiny as the demographic biases that have dominated the algorithmic
hiring literature.

\section{Methods}\label{sec3}

To study generative preference bias in the within-AI setting, we need
a framework that (a) controls for candidate quality, so that any
observed differences in hiring outcomes can be attributed to model
choice, (b) separates the effects of writer model
from evaluator model, and (c) is replicable as the model landscape
changes. We design our framework around these requirements, operating
as the three-phase pipeline summarised in Figure~\ref{fig:pipeline}
(generation, evaluation, analysis); what follows fills in the
implementation detail behind each phase. We
use two publicly available datasets, one of CVs and one of job
descriptions \cite{resume_kaggle, jobDesc_kaggle}, and select ten job
descriptions that range widely in industry and seniority, from Primary
School Teacher to Chief Financial Officer. For each job we curate a pool of fifty candidate
CVs by computing cosine similarity between CV embeddings and the job
description using the all-MiniLM-L6-v2 sentence
encoder \cite{reimers_sentence-bert_2019}. The top 25 candidates by
similarity are designated \emph{high-fit} candidates; the remaining 25
are \emph{moderate-fit} candidates (ranks 26--50 within the pool),
drawn from a middle tier of the distribution.
This is a deliberate design choice: early test runs showed that LLM
evaluators trivially filter out irrelevant profiles. Using candidates
with some relevant background instead ensures the experiment tests
nuanced distinctions; a bottom-tier pool would only test the capacity
to reject obvious noise.

Cosine similarity between a CV and a job description is a coarse proxy
for occupational fit — it can be thrown off by superficial vocabulary
overlap or miss substantive but differently-worded qualifications — but
it has a property none of the nine evaluator models can offer on their
own: it is a single, external, evaluator-independent criterion, so
every evaluator is judged against the exact same high-fit/moderate-fit
boundary (each evaluator's own
CV-only judgment would draw that line somewhat differently on its own — by 6\% to
14\% of candidates, Table~\ref{tab:cosine_agreement} in
Appendix~\ref{secA}). A fixed, model-agnostic criterion is what makes
the cross-evaluator comparisons throughout the rest of the paper
meaningful: ``high-fit'' refers to the same fifty candidates regardless
of whose scores are being read. Despite being coarse, it also turns out
to track what evaluators themselves perceive as quality well:
Appendix~\ref{secA4} shows a large, consistent CV-only scoring
gap between the two tiers ($\Delta = 2.25$ on a 1--10 scale), the
highest inter-evaluator agreement of any condition ($\bar{\rho} =
0.92$), and substantial agreement between the cosine-similarity split
itself and each evaluator's own independent CV-only ranking (86--94\%
of candidates assigned to the same tier either way, per evaluator;
Table~\ref{tab:cosine_agreement}) — evidence that the stratification
captures a real, substantial, and independently-verified quality
difference.

The one analysis that deliberately departs from this fixed criterion is
the rank-displacement analysis of Section~\ref{sec:competitiveness},
which asks a self-referential question — does cover-letter quality
overturn an evaluator's \emph{own} internal ranking of candidates? —
and so needs each evaluator's own CV-only ranking as its baseline. The
shared cosine-similarity split would not work here: using an external
criterion there would conflate genuine writer-driven displacement with
an evaluator simply disagreeing with that external criterion to begin
with.

The \textbf{generation phase}'s writer pool comprises nine API-hosted models
(GPT-4o-mini, GPT-5-mini, and GPT-5 by OpenAI; Claude~Haiku~4.5 and
Claude~Sonnet~4.6 by Anthropic; Gemini~3~Flash and Gemini~3.5~Flash by
Google; DeepSeek~V3.2 and DeepSeek~V4~Flash by DeepSeek) and two locally
hosted models (Llama-3.1-8B by Meta; DeepSeek-R1-8B by
DeepSeek).\footnote{DeepSeek's API-hosted models (V3.2, V4 Flash) are
themselves open-weight, publicly released alongside the proprietary
offerings from OpenAI, Anthropic, and Google that make up the rest of
the API-hosted pool. We therefore classify writers and evaluators
throughout by \emph{access modality} (API-hosted vs.\ locally hosted).
Access modality, and the cost and infrastructure burden it implies, is
the axis our framework is designed to measure; weight availability and
licensing are incidental to that classification. DeepSeek-R1-8B, the one
DeepSeek model we host locally, is small enough to run on consumer
hardware; DeepSeek's larger, open-weight models are, like the
proprietary offerings, in practice accessed only via API.} The API-hosted pool spans both cost tiers available to
employers: lightweight, cost-efficient ``mini''/``flash'' variants and
full flagship models (GPT-5, Claude Sonnet 4.6), while applicants
without access to paid APIs may resort to locally hosted models running
at no cost. Including both access tiers allows us to directly measure
the competitive consequences of this accessibility gap. Model size and
hosting modality are consequently confounded by design. Our two locally
hosted models are both around 8B parameters, chosen to match roughly
the ceiling of what runs on consumer hardware, the same constraint that
defines the applicant population this tier is meant to represent, and
not to compare model scale for its own sake. A locally hosted model
large enough to match the flagship API models would need cluster-scale
compute well beyond an unpaid applicant's reach; at that point the
applicant is again dependent on paid infrastructure access, just of a
different kind than an API subscription. Isolating a pure model-size
effect, independent of hosting cost, would require a same-size,
different-hosting comparison, which we leave to future work. All models use
provider default temperature and sampling parameters, so that each
model operates under its own default regime as it would in real-world
deployment. No output-length limit is
set explicitly for any model, with one exception: Anthropic's Messages
API requires \texttt{max\_tokens} as a mandatory field, so Claude
models alone receive an explicit value (4,096
during generation, 1,024 during evaluation), well above the
350--400 word target described next, so it does not bind in practice.
Each writer is
instructed to produce persuasive, 350--400 word letters without
fabricating credentials; the word limit is enforced at the prompt level
to suppress length-based quality inflation
\cite{saito_verbosity_2023}. A fixed gender-neutral signoff is used
throughout to prevent evaluators from inferring applicant gender from
the name. The full prompt template is reproduced verbatim in
Appendix~\ref{secD}.

The \textbf{evaluation phase}'s nine API-hosted evaluator models (the
same models used as writers, excluding the two locally hosted ones)
score each candidate on a 1--10 scale with explicit verbal descriptors
for each range: scores 1--3 indicate a completely unqualified candidate,
4--6 indicate relevant background but insufficient specific experience,
7--8 indicate a strong match, and 9--10 indicate a perfect fit. This
rubric is held constant across all evaluator models and conditions to
standardize scoring behavior. Across the three conditions introduced in
Results (Section~\ref{sec4}), the only element that changes is which
documents are provided; all other prompt content is held identical.
Scores are extracted via XML-style delimiters
(\texttt{<final>}~\ldots~\texttt{</final>}), with anything outside the
tags discarded; this prevents chain-of-thought leakage from affecting
extracted scores, which is especially relevant for reasoning-capable
models such as DeepSeek-R1. Ties are broken merit-first: a moderate-fit
candidate can only displace a high-fit candidate by achieving a
\emph{strictly} higher score.

\subsection{Writing-Style Feature Definitions}\label{sec:methods_features}

Appendix~\ref{secA1} characterises writer output
along five feature classes, measured as follows. Word count, sentence
count, paragraph count, and comma count are literal surface counts
extracted from the letter text. Average word length (characters per
word) and type-token ratio (TTR, capturing the fraction of words in the
letter that are unique: 0 means every word appears at least twice while
1 means none do) proxy vocabulary richness; Flesch Reading Ease scores
readability on a scale where higher means easier to read (roughly:
90--100 simple text, 60--70 plain English understood by most adults,
30--50 fairly difficult/academic, 0--30 very difficult/dense; the score
can dip slightly negative for extremely dense text). VADER compound
score ranges from $-1$ (most negative) to $+1$ (most positive). VAD
Valence, Arousal, and Dominance (NRC-VAD lexicon) each range from $-1$
to $+1$ with $0$ as the neutral midpoint, scoring tone (negative
$\leftrightarrow$ positive), energy (calm $\leftrightarrow$ excited),
and control (submissive $\leftrightarrow$ dominant) respectively. The
emotion classifier outputs a probability over seven categories (Joy,
Neutral, Surprise, Fear, Anger, Disgust, Sadness) that sum to $1$ for
every letter by construction. Cosine similarity is computed between the
letter's embedding and the job-ad embedding, and separately between the
letter's embedding and the CV embedding; both range from $-1$ to $1$ in
principle, though in practice professional-text pairs score positively.

\subsection{Calibration Metrics}\label{sec:methods_calibration}

Appendix~\ref{secA4} relies on three measurements,
defined here. Each (job $j$, candidate $c$, writer $w$, evaluator $e$)
combination was scored in four independent API calls with no shared
context between runs, yielding four scores $s^{(r)}_{e,j,c,w}$ for run
$r \in \{1,2,3,4\}$. Reproducibility is measured as the within-group
standard deviation $\sigma_{e,j,c,w} =
\mathrm{std}(\{s^{(r)}_{e,j,c,w}\}_{r=1}^{4})$, reported as a mean per
evaluator and condition. Given the high reproducibility documented in
Appendix~\ref{secA4}, the rest of the paper associates each
(job, candidate, writer, evaluator) tuple with a single score
$s_{e,j,c,w}$, defined as the mean over the four independent runs. The
tier discrimination gap is $\Delta = \bar{s}_{\mathrm{high}} -
\bar{s}_{\mathrm{mod}}$, the difference between the mean score of
high-fit and moderate-fit candidates for a given evaluator and
condition. Inter-evaluator agreement on candidate quality is measured
with Spearman's rank correlation $\rho$, which is invariant to monotone
rescaling and therefore robust to strictness differences across
evaluators: for each pair of evaluators, $\rho$ is computed over their
score vectors across all (job, candidate) pairs, ranking the 50
candidates per job and correlating those rankings between evaluators,
then aggregating across jobs. In the CV-only condition this is done
directly over $s_{e,j,c}$; in the cover-letter conditions, separately
for each writer model over $s_{e,j,c,w}$, then averaged across the
eleven writers.

\FloatBarrier

\section*{Acknowledgements}

This work was supported as a part of NCCR Automation, a National Centre
of Competence in Research, funded by the Swiss National Science
Foundation (grant number 51NF40\_225155).

\section*{Declarations}

\noindent\textbf{Funding.} This work was supported as a part of NCCR
Automation, a National Centre of Competence in Research, funded by the
Swiss National Science Foundation (grant number 51NF40\_225155).

\noindent\textbf{Conflict of interest/Competing interests.} The authors
declare no competing interests.

\noindent\textbf{Ethics approval and consent to participate.} Not
applicable: this study did not involve human participants, human data,
or animal subjects. All CVs and job descriptions used are drawn from
publicly available datasets \cite{resume_kaggle, jobDesc_kaggle} and
were anonymized prior to analysis.

\noindent\textbf{Consent for publication.} Not applicable.

\noindent\textbf{Data availability.} The base CV and job description
datasets are publicly available \cite{resume_kaggle, jobDesc_kaggle}.
The derived experimental data (generated cover letters and evaluator
scores) are available from the corresponding author upon reasonable
request.

\noindent\textbf{Materials availability.} Not applicable.

\noindent\textbf{Code availability.} The auditing framework code is
available from the corresponding author upon reasonable request.

\noindent\textbf{Author contribution.} All authors contributed to the
conception and design of the auditing framework. All authors
contributed to data analysis and interpretation, and to drafting and
revising the manuscript. All authors read and approved the final
manuscript.

\bibliographystyle{unsrtnat}
\bibliography{references}

\newpage

\begin{appendices}

\makeatletter
\@addtoreset{figure}{section}
\@addtoreset{table}{section}
\@addtoreset{equation}{section}
\makeatother
\renewcommand{\thefigure}{\thesection\arabic{figure}}
\renewcommand{\thetable}{\thesection\arabic{table}}
\renewcommand{\theequation}{\thesection\arabic{equation}}
\renewcommand{\theHfigure}{app.\thefigure}
\renewcommand{\theHtable}{app.\thetable}
\renewcommand{\theHequation}{app.\theequation}

\section{Validating the Experimental Design}\label{secA}

This appendix gives the full detail behind the two preliminaries from
the start of the Results section (Section~\ref{sec:prelim}). For the
writer-genuineness
preliminary: the complete per-model writer text-feature breakdown
(\ref{secA1}), and two targeted sanity checks for whether writers'
outputs vary systematically along dimensions the design depends on:
job identity (\ref{secA2}) and candidate tier (\ref{secA3}). For the
evaluator-calibration preliminary: the complete evaluator-calibration
breakdown (\ref{secA4}), and a third sanity check, whether the fixed
cosine-similarity tier split agrees with what evaluators themselves
would conclude (\ref{secA5}). Last, a check that the generative
preference hierarchy of Section~\ref{sec:preference_hierarchy} is a
real cross-evaluator signal and not noise: how often evaluators agree
on which writer wrote the best letter for the same candidate
(\ref{secA6}).

\subsection{Writer Model Text-Feature Results}\label{secA1}

\begin{figure}[!htbp]
  \centering
  \includegraphics[width=\textwidth]{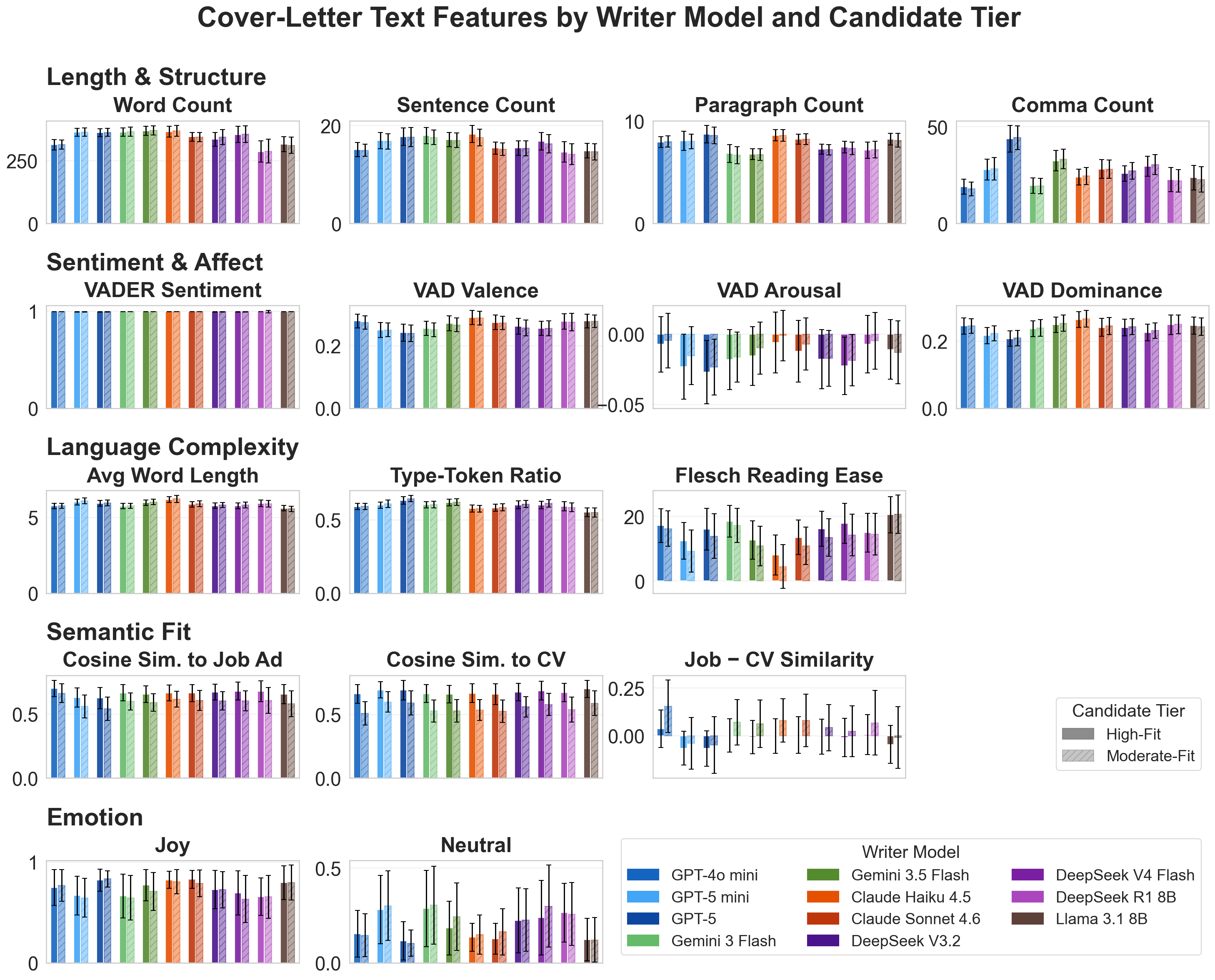}
  \caption{Cover-letter text features by writer model and candidate
    tier, grouped by feature class (one row per class). Emotion is
    reduced to Joy and Neutral, the two dominant categories; the
    remaining five (surprise, fear, anger, disgust, sadness) are low, largely indistinguishable across models, and, thus, omitted here.
    Error bars show $\pm 1$ standard deviation across candidates. The
    three Semantic Fit columns verify the alignment of the \emph{generated cover
    letter's} embedding with the ones of the job ad and CV: ``Cosine Similarity to Job Ad'' is
    letter-vs-job-ad similarity, ``Cosine Similarity to CV'' is
    letter-vs-CV similarity, and ``Job $-$ CV Similarity'' is their
    difference, i.e.\ whether a given letter leans more toward echoing
    the job posting's language or the candidate's own CV.}
  \label{fig:writer_features}
\end{figure}

This section gives the full per-model detail behind the summary at the
start of the Results section (Section~\ref{sec4}). Figure~\ref{fig:writer_features}
plots each feature's mean and standard deviation by writer model and
candidate tier. The five paragraphs below go through the feature
classes one at a time; the two sanity checks mentioned in the main
text are detailed separately in \ref{secA2} and \ref{secA3}.

\paragraph{Length and structure.}
Word count, sentence count, paragraph count, and comma count are literal
surface counts extracted from the letter text. Mean word count ranges
from 288 words (DeepSeek-R1-8B) to 370 words (Gemini~3.5~Flash), with
most models clustering in the 340--370 range, close to the 350--400
word target given in the prompt. Sentence count spans a narrower range,
14.4 (DeepSeek-R1-8B) to 17.9 (Claude~Haiku~4.5) sentences per letter.
Paragraph count separates the two Gemini models (6.75 paragraphs, the
fewest of any writer) from GPT-5 and Claude~Haiku~4.5 (8.6--8.7, the
most): the Gemini models favour fewer, longer paragraphs, where the
others break similar content into more, shorter ones. Comma count is
the one clear outlier: GPT-5 averages 44.1 commas per letter, close to
double the next-highest writer (DeepSeek~V4~Flash, 30.0) and more than
double the lowest (GPT-4o-mini, 18.4): substantially more
syntactically complex, clause-heavy sentences than any other model.
Verbosity and structure are stable per-model traits, quantified
precisely in the tier-adaptation sanity check below: they reflect a
fixed model style.

\paragraph{Language complexity.}
Average word length (characters per word) and type-token ratio (TTR, capturing the fraction of words in the letter that are unique: 0 means every word
appears at least twice while 1 means none do) proxy vocabulary richness; Flesch Reading
Ease scores readability on a scale where higher means easier to read
(roughly: 90--100 simple text, 60--70 plain English understood by most
adults, 30--50 fairly difficult/academic, 0--30 very difficult/dense;
the score can dip slightly negative for extremely dense text). All
eleven writers fall in the 6--21 range on Flesch: dense, formal,
professional prose throughout, well below the ``plain English'' band.
Within that narrow band, Claude~Haiku~4.5 writes both the longest
average words (6.20 characters) and the least readable letters (Flesch
$=6.2$); Llama-3.1-8B writes the shortest average words (5.59
characters) and the most readable letters of the eleven (Flesch
$=20.6$, itself still within the ``very difficult'' band). Vocabulary
richness varies separately from either of these: GPT-5 has the highest
TTR (0.638, the least word repetition) and Llama-3.1-8B the lowest
(0.550).

\paragraph{Sentiment and affect.}
VADER compound score ranges from $-1$ (most negative) to $+1$ (most
positive). VAD Valence, Arousal, and Dominance (NRC-VAD lexicon) each
range from $-1$ to $+1$ with $0$ as the neutral midpoint, scoring tone
(negative $\leftrightarrow$ positive), energy (calm $\leftrightarrow$
excited), and control (submissive $\leftrightarrow$ dominant)
respectively. VADER sits essentially at its ceiling for every writer
(mean $0.996$; writer means span only $0.0021$), confirming letters are
uniformly and almost maximally positive in polarity: this feature does
not discriminate between writers at all. VAD Valence is positive on
average ($0.266$) but far short of its $+1$ ceiling, with modest spread
across writers ($0.241$--$0.289$): a mildly positive tone throughout.
VAD Arousal sits close to the neutral midpoint for every writer (all
means between $-0.025$ and $-0.004$) and is the tightest of the three
VAD axes across writers; VAD Dominance is the widest ($0.209$--$0.266$),
meaning writers differ more in how assertive their letters read than in
how positive or energetic they are. GPT-5 scores lowest of all eleven
writers on all three VAD axes; Claude~Haiku~4.5 scores highest on all
three.

\paragraph{Emotion.}
The emotion classifier outputs a probability over seven categories
(Joy, Neutral, Surprise, Fear, Anger, Disgust, Sadness) that sum to $1$
for every letter by construction. Joy and Neutral together account for
the large majority of that probability mass; the other five combined
average only $6\%$ across all letters (at most $9\%$ for any single
writer). Writers mainly differ in how
they split the Joy/Neutral majority: GPT-5 is the most Joy-leaning
($0.826$ Joy, $0.110$ Neutral), with Claude~Haiku~4.5 and
Claude~Sonnet~4.6 close behind ($\approx 0.81$ Joy); Gemini~3~Flash is
the most Neutral-leaning ($0.654$ Joy, $0.296$ Neutral), with
GPT-5-mini and DeepSeek-R1-8B similarly measured ($\approx 0.66$ Joy,
$0.26$--$0.29$ Neutral). Every writer is predominantly Joy-toned, though
they differ in how strongly they favor Joy over Neutral.

\paragraph{Semantic fit to the job description.}
Cosine similarity is computed between the letter's embedding and the
job-ad embedding, and separately between the letter's embedding and the
CV embedding; both range from $-1$ to $1$ in principle, though in
practice professional-text pairs score positively. The clearest pattern
here is a \emph{tier} effect: for High-Fit
candidates, the letter is on average very slightly more similar to the
CV than to the job ad (job$-$CV $=-0.013$); for Moderate-Fit candidates,
that flips to a real lean toward the job ad instead (job$-$CV
$=+0.045$). This shift holds broadly: every one of the eleven models
shows it (Figure~\ref{fig:writer_features},
bottom-right panel). A plausible reading: when the underlying CV is a
weaker match for the job, as Moderate-Fit candidates are by
construction, the letter compensates by echoing the job description's
own language more than it reflects the CV's actual, weaker content. On top of this shared tier effect, writers also differ
in their overall baseline: pooled across both tiers, GPT-4o-mini leans
furthest toward the job ad (job$-$CV $=+0.096$), while GPT-5,
GPT-5-mini, and Llama-3.1-8B lean toward the CV instead ($-0.03$ to
$-0.06$).

\subsection{Content Consistency Across Jobs}\label{secA2}

\begin{figure}[!htbp]
  \centering
  \includegraphics[width=0.85\textwidth]{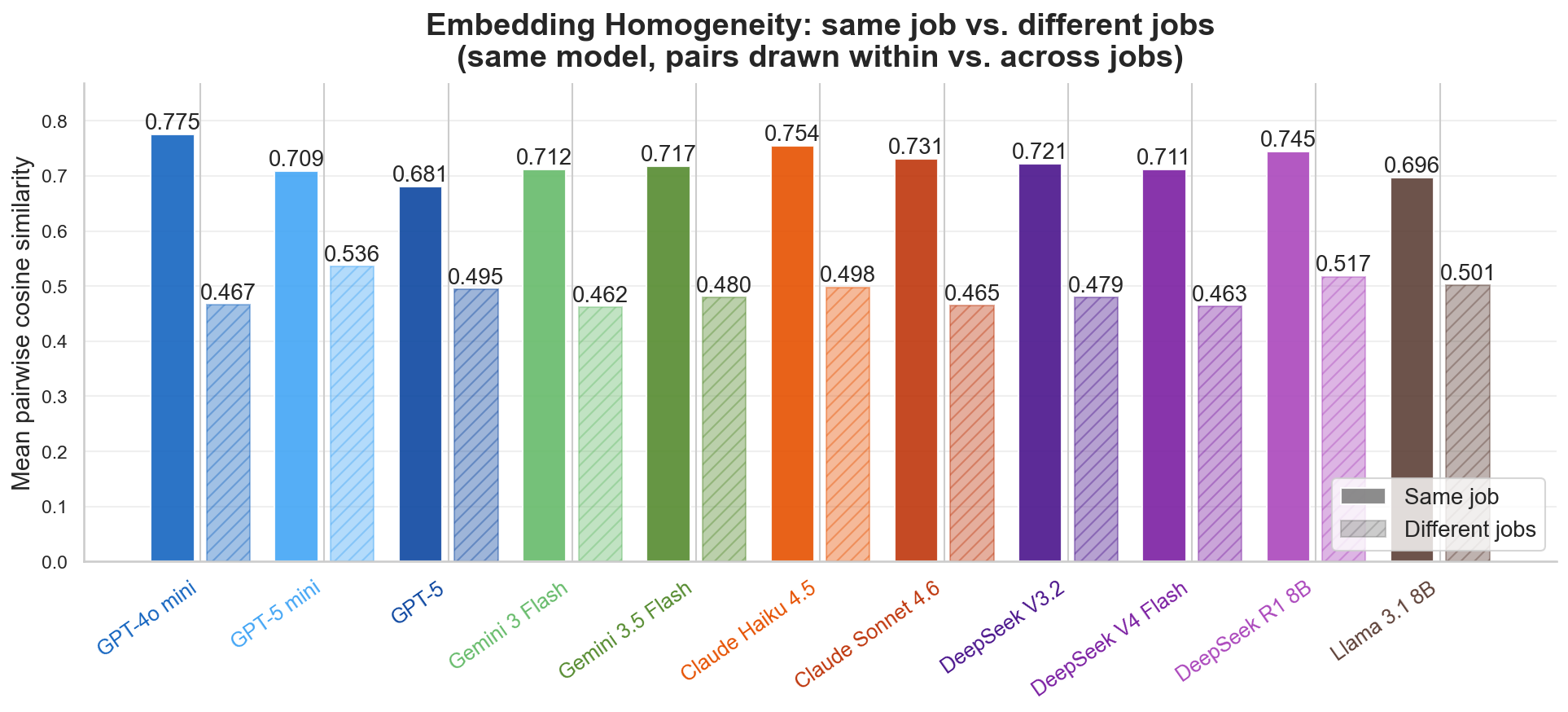}
  \caption{Mean pairwise cosine similarity between cover-letter
    embeddings for the same writer, comparing pairs drawn from the same
    job vs.\ different jobs. Within-job similarity exceeds cross-job
    similarity for every writer, confirming that writers condition
    their output on the specific job description. This validates the
    experimental design
    (mentioned at the start of the Results section, Section~\ref{sec4}).}
  \label{fig:homogeneity}
\end{figure}

Whether a writer's output for a given job is recognisably tied to that
job, or largely interchangeable regardless of which job it was written
for. Figure~\ref{fig:homogeneity} compares, for each writer, mean
pairwise cosine similarity between letter embeddings drawn from the
same job against pairs drawn from different jobs.

Within-job similarity (0.68--0.78) is far higher than cross-job
similarity (0.46--0.54) for every writer. This validates the
experimental setup: writers genuinely condition the letter's content
on the specific job description, which is what the auditing framework
requires for job-to-job and writer-to-writer comparisons to be
meaningful in the first place. GPT-5-mini has the smallest
within/cross-job gap (its output varies least from job to job), while
GPT-5 has the lowest within-job similarity (it preserves the most
letter-to-letter individuality even among candidates applying to the
same job).

\subsection{Style Adaptation to Candidate Tier}\label{secA3}

\begin{figure}[!htbp]
  \centering
  \includegraphics[width=0.92\textwidth]{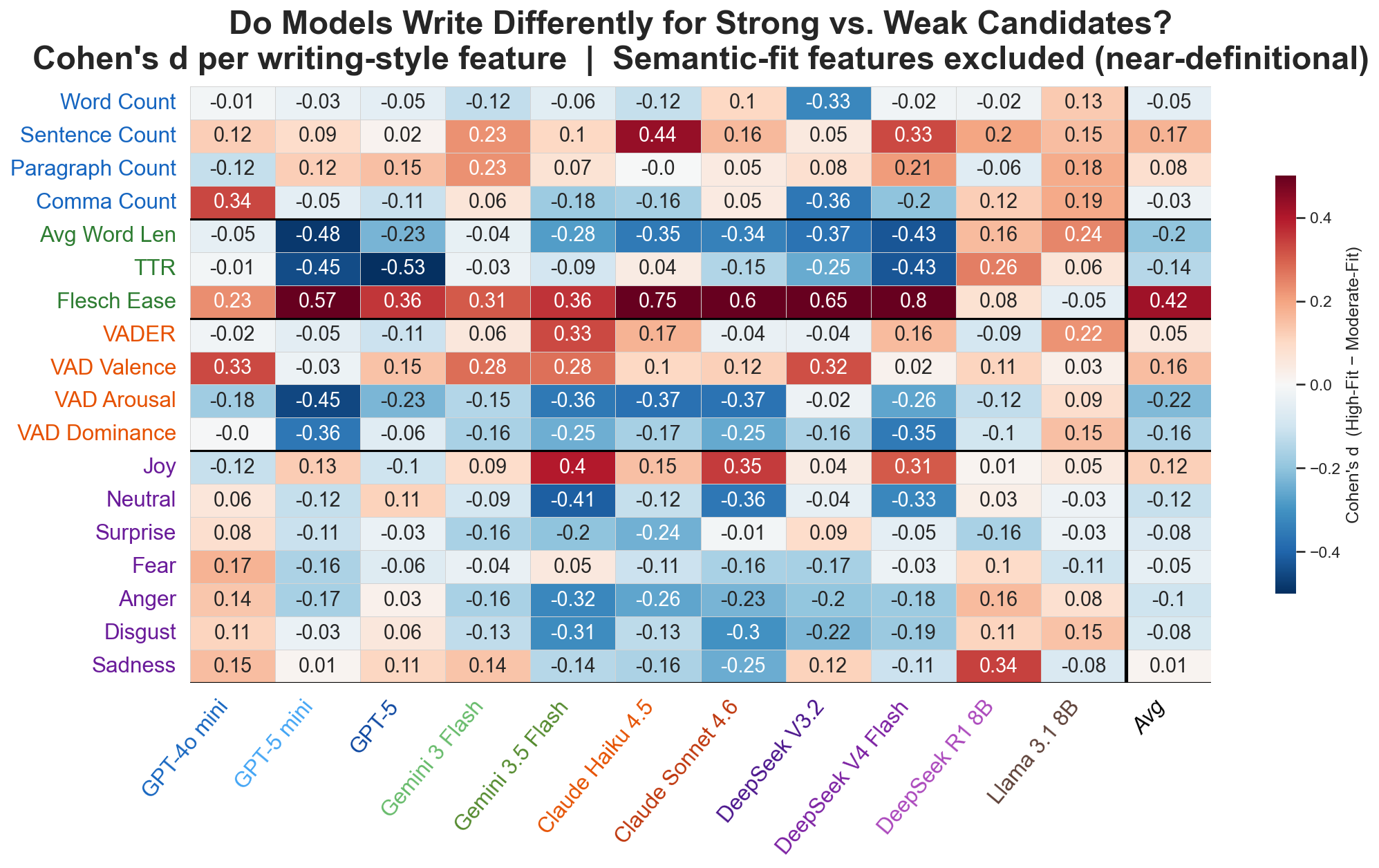}
  \caption{Cohen's $d$ (High-Fit $-$ Moderate-Fit) per writing-style
    feature and writer model, averaged across jobs. Flesch Reading Ease
    is the one systematically positive effect across almost every
    writer; other features show smaller and less consistent effects
    (summarised at the start of the Results section, Section~\ref{sec4}).}
  \label{fig:tier_effect}
\end{figure}

Whether writers adapt \emph{how} they write, beyond simply what they
write about, when the underlying candidate is a stronger or
weaker match for the job. Figure~\ref{fig:tier_effect} reports Cohen's
$d$ (High-Fit $-$ Moderate-Fit) for every writing-style feature and
writer, averaged across jobs; semantic-fit features are excluded here
since their tier effect is near-definitional (a High-Fit CV is, by
construction, closer to the job, so a letter echoing it will be too).

One feature stands out: Flesch Reading Ease is consistently higher for
High-Fit candidates across almost every writer, up to $d = +0.80$ for
DeepSeek~V4~Flash, i.e.\ letters for stronger candidates are written in
simpler, more accessible language. GPT-5 ($d = 0.08$) and Llama-3.1-8B
($d = -0.05$) barely adapt to tier at all, while the DeepSeek and Claude
models adapt the most, primarily via readability. A fainter
mirror-image pattern appears in average word length, type-token ratio,
VAD Arousal, and VAD Dominance, which skew slightly negative for
High-Fit candidates on average ($\bar d = -0.20, -0.14, -0.22, -0.16$
respectively); emotion features show no consistent direction. Taken
together, this multi-feature shift mirrors the semantic-fit
compensation described in \ref{secA1}: when the underlying case for a
candidate
is less obvious, writers reach for longer, rarer words and a more
assertive, more energetic register, as though compensating for a
weaker fit with more elaborate language.

\subsection{Evaluator Calibration Details}\label{secA4}

Full per-evaluator detail behind the summary at the start of the
Results section (Section~\ref{sec4}). Reproducibility ($\sigma$), the tier
discrimination gap ($\Delta$), and the inter-evaluator agreement
procedure ($\rho$) are defined in Methods
(Section~\ref{sec:methods_calibration}).
\begin{table}[h]
\centering
\caption{Mean intra-run $\sigma$ by evaluator and condition. Evaluators are
  ordered left-to-right by increasing mean $\sigma$ (most to least
  reproducible), using the same colour-coding as elsewhere in the paper.
  Scores on a 1--10 scale, rounded to two decimals. CV = CV
  only; CL = cover letter only; CV+CL = CV + cover letter.}
\label{tab:intra_run_variance}
\small
\setlength{\tabcolsep}{4pt}
\begin{tabular}{@{}l*{9}{c}|c@{}}
\toprule
 & {\footnotesize\textcolor[HTML]{BF360C}{\shortstack[t]{Claude\\Sonnet 4.6}}}
 & {\footnotesize\raisebox{-1.3em}{\textcolor[HTML]{E65100}{\shortstack[t]{Claude\\Haiku 4.5}}}}
 & {\footnotesize\textcolor[HTML]{0D47A1}{\shortstack[t]{GPT-5\\\ }}}
 & {\footnotesize\raisebox{-1.3em}{\textcolor[HTML]{558B2F}{\shortstack[t]{Gemini 3.5\\Flash}}}}
 & {\footnotesize\textcolor[HTML]{1565C0}{\shortstack[t]{GPT-4o\\mini}}}
 & {\footnotesize\raisebox{-1.3em}{\textcolor[HTML]{66BB6A}{\shortstack[t]{Gemini 3\\Flash}}}}
 & {\footnotesize\textcolor[HTML]{42A5F5}{\shortstack[t]{GPT-5\\mini}}}
 & {\footnotesize\raisebox{-1.3em}{\textcolor[HTML]{7B1FA2}{\shortstack[t]{DeepSeek V4\\Flash}}}}
 & {\footnotesize\textcolor[HTML]{4A148C}{\shortstack[t]{DeepSeek\\V3.2}}}
 & {\shortstack[t]{\textbf{\textit{Mean}}\\\ }} \\
\midrule
\textbf{CV}    & 0.04 & 0.13 & 0.21 & 0.23 & 0.29 & 0.23 & 0.36 & 0.37 & 0.40 & \textbf{\textit{0.25}} \\
\textbf{CL}    & 0.07 & 0.10 & 0.30 & 0.34 & 0.30 & 0.38 & 0.39 & 0.55 & 0.62 & \textbf{\textit{0.34}} \\
\textbf{CV+CL} & 0.06 & 0.11 & 0.23 & 0.28 & 0.26 & 0.30 & 0.40 & 0.49 & 0.48 & \textbf{\textit{0.29}} \\
\midrule
\textbf{Avg}   & 0.05 & 0.11 & 0.25 & 0.28 & 0.29 & 0.30 & 0.38 & 0.47 & 0.50 & \textbf{\textit{0.29}} \\
\bottomrule
\end{tabular}
\end{table}

\begin{figure}[!htbp]
  \centering
  \begin{minipage}[t]{\textwidth}
    \centering
    \includegraphics[width=0.95\textwidth]{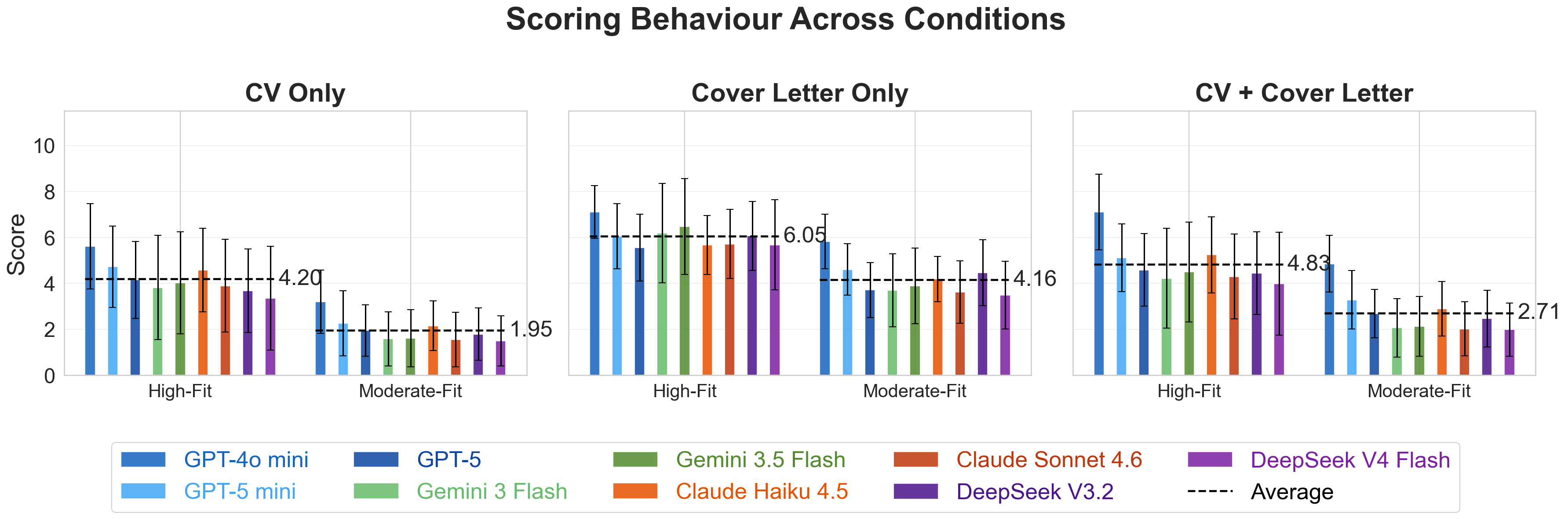}
    \par\smallskip
    \textbf{(a)}
  \end{minipage}
  \par\medskip
  \begin{minipage}[t]{\textwidth}
    \centering
    \includegraphics[width=0.95\textwidth]{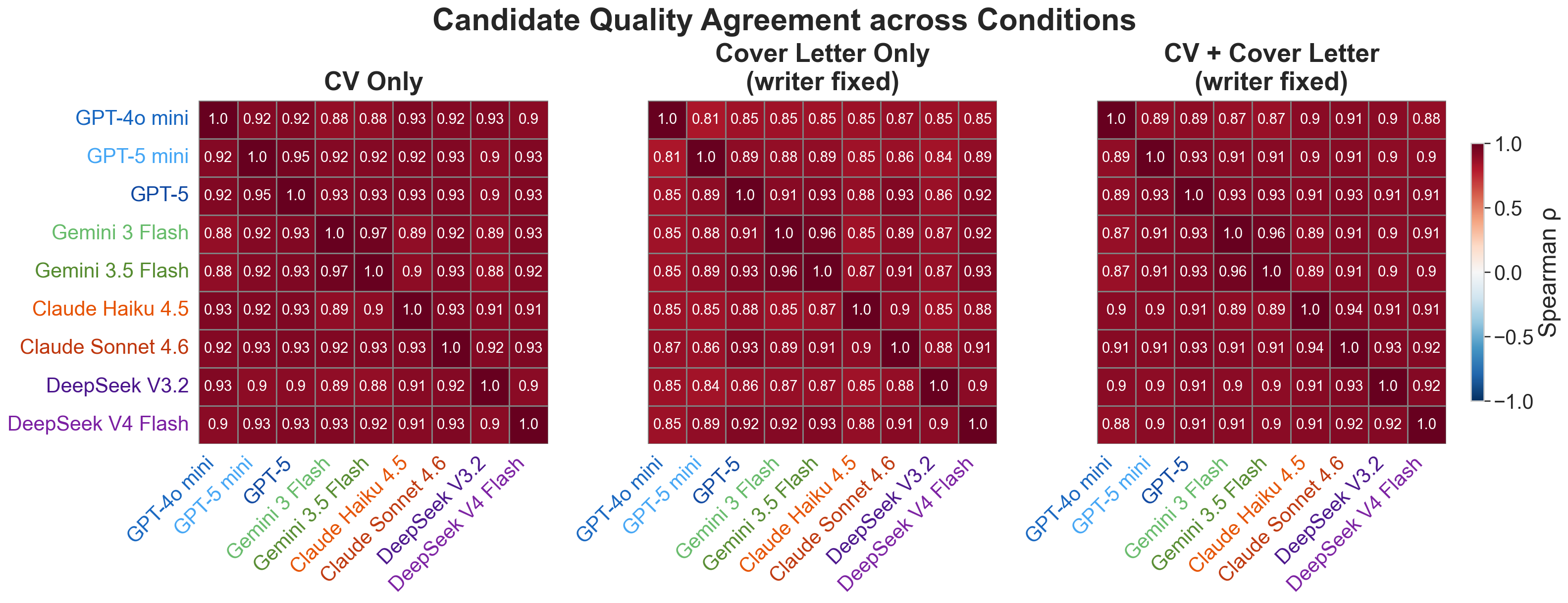}
    \par\smallskip
    \textbf{(b)}
  \end{minipage}
  \caption{\textbf{(a)} Mean evaluator scores ($\pm 1$ standard
    deviation) for high-fit and moderate-fit candidates across the three
    evaluation conditions. \textbf{Left (CV only):} all evaluators
    discriminate in the expected direction (overall gap $\Delta = 2.25$),
    confirming that the stratification is universally recognized.
    \textbf{Centre (cover letter only):} scores inflate markedly and the
    gap narrows ($\Delta = 1.89$), as AI-generated prose presents all
    candidates favourably regardless of merit. \textbf{Right (CV +
    cover letter):} combining documents partially restores discrimination
    ($\Delta = 2.12$).
    \textbf{(b)} Candidate quality agreement: Spearman $\rho$ between
    evaluator score vectors across all conditions. \textbf{Left:}
    CV-only ($\bar{\rho} = 0.92$). \textbf{Centre:} cover-letter-only
    ($\bar{\rho} = 0.88$). \textbf{Right:} combined CV + Cover letter
    ($\bar{\rho} = 0.91$). Despite strictness differences, evaluators
    agree strongly on candidate quality across all conditions.}
  \label{fig:strictness_by_condition}
\end{figure}

Claude~Sonnet~4.6 and Claude~Haiku~4.5 are the most stable; DeepSeek~V3.2
the most variable, peaking at $\sigma = 0.623$ in the cover-letter-only
condition. Given that scores are integers, even this upper end
corresponds to sub-single-point fluctuations on average across runs, so
scoring remains broadly consistent. Additional analysis confirms that
writer identity has little effect on variability: the dominant source is
the evaluator model.

\paragraph{Scoring behavior across conditions.}
With reproducibility established, we turn to how evaluators score
candidates under different conditions. Figure~\ref{fig:strictness_by_condition}(a)
shows mean evaluator scores for high-fit and moderate-fit candidates
across all three conditions. We first focus on the CV-only condition,
which isolates evaluator behavior from any generative influence: no
cover letter is involved, so scores reflect only the evaluators'
intrinsic response to the raw candidate profiles. Throughout, we measure the tier discrimination gap as $\Delta =
\bar{s}_{\mathrm{high}} - \bar{s}_{\mathrm{mod}}$, the difference
between the mean score of high-fit and moderate-fit candidates for a
given evaluator and condition. In the CV-only condition, high-fit
candidates receive an overall mean of 4.20 and moderate-fit candidates
1.95, giving $\Delta = 2.25$ averaged across evaluators. This separation holds across all models,
from the largest gap (GPT-5-mini: $\Delta = 2.47$) to the smallest
(DeepSeek~V4~Flash: $\Delta = 1.87$), confirming that the
cosine-similarity stratification created two distinctly tiered groups
that LLM evaluators universally recognize, despite the deliberately
conservative top-half/bottom-half split described in
Methods (Section~\ref{sec3}): the two tiers differ substantially in
evaluator-perceived quality, beyond the embedding similarity that
defines them. This
CV-only condition also has the highest inter-evaluator agreement of the
three conditions ($\bar{\rho} = 0.92$, below), which is itself
corroborating evidence that the tier split tracks a real quality
signal: independent
evaluators, seeing only the raw CV, converge almost perfectly on which
tier a candidate belongs to. For a more direct check, we compared the
cosine-similarity split against the high-fit/moderate-fit split each
evaluator's own CV-only scores would draw on their own. The two agree
on about 90\% of candidates on average, from 94\% for
Claude~Sonnet~4.6 (the closest match) down to 86\% for
DeepSeek~V3.2 (the loosest); no evaluator falls far outside that range
(full breakdown in Table~\ref{tab:cosine_agreement}, Appendix~\ref{secA}).
The two splits are not identical: about one candidate in ten lands in
a different tier under an evaluator's own judgment, for every evaluator.
This makes a more direct
case that the cosine-similarity split tracks real quality than the
aggregate $\Delta$ and inter-evaluator $\rho$ reported above. Standard deviations exhibit
a notable asymmetry: variability is consistently higher for high-fit
candidates ($\sigma \approx 1.7$--$2.3$) than for moderate-fit
candidates ($\sigma \approx 1.1$--$1.4$), suggesting that evaluators
apply more heterogeneous judgements to strong CVs where subtle quality
differences are harder to resolve. Models do, however, differ
substantially in absolute strictness: GPT-4o-mini assigns the highest
scores overall ($\mu = 4.40$) while DeepSeek~V4~Flash is the most
conservative ($\mu = 2.42$).

The tier discrimination persists across all three conditions, with some
variation. When evaluators see only cover letters, scores inflate
markedly for both tiers (overall means: 6.05 vs.\ 4.16) and the gap
narrows to $\Delta = 1.89$, as AI-generated prose presents all
candidates favourably regardless of merit. Two evaluators stand out as
outliers in opposite directions: Gemini~3.5~Flash maintains the largest
gap of any model ($2.59$) and scores moderate-fit candidates
considerably lower than its peers ($3.89$ vs.\ a cross-model average of
$4.16$), while GPT-4o-mini maintains the smallest gap ($1.28$),
extending its overall leniency into the moderate-fit tier ($5.82$).
Combining CV and cover letter partially restores discrimination
($\Delta = 2.12$), with score levels settling between the two
single-document conditions (high-fit: 4.83; moderate-fit: 2.71).
GPT-4o-mini again assigns the highest scores to high-fit candidates
($7.10$), while Gemini~3.5~Flash produces the widest spread ($2.36$) and
DeepSeek~V4~Flash the lowest moderate-fit scores ($1.99$). These
inter-model differences in absolute strictness, consistent across
conditions, motivate the use of rank-based, strictness-neutral metrics
throughout.

\paragraph{Candidate quality agreement.}
Having characterised individual scoring behavior, we now ask whether
evaluators reach a shared verdict on candidate quality despite their
different strictness levels. We measure inter-evaluator agreement using
Spearman's rank correlation $\rho$, which is invariant to monotone
rescaling and therefore robust to the strictness differences documented
above. For each pair of evaluators, $\rho$ is computed over their score
vectors across all (job, candidate) pairs: for each job, the 50
candidates are ranked by score, and $\rho$ is computed between
evaluators over these rankings, then aggregated across jobs. In the
CV-only condition this is done directly over $s_{e,j,c}$; in the
cover-letter conditions, separately for each writer model over
$s_{e,j,c,w}$, then averaged across the eleven writers. The resulting
matrices (Figure~\ref{fig:strictness_by_condition}(b)) show high
agreement across all three conditions: $\bar{\rho} = 0.92$ (range
0.88--0.97) in the CV-only condition, $\bar{\rho} = 0.88$ (range
0.81--0.96) in the
cover-letter-only condition, and $\bar{\rho} = 0.91$ (range 0.87--0.96) in the combined
condition. The small standard deviations of the per-writer correlations
(shown per cell) confirm that this agreement holds regardless of which
writing style is used. Since the writer is held fixed within each
correlation, the agreement reflects evaluator consensus on candidate
quality alone, confirming that the evaluators are calibrated enough for
the writer-preference analysis that follows to be meaningful.

\subsection{Cosine-Similarity Tier Split vs.\ Evaluator CV-Only Judgment}\label{secA5}

Methods (Section~\ref{sec3}) uses a single, fixed cosine-similarity split to
define the high-fit/moderate-fit tiers, precisely so that ``high-fit''
means the same fifty candidates for every evaluator. Table~\ref{tab:cosine_agreement}
quantifies how much this fixed split actually costs relative to asking
each evaluator directly: for every evaluator, we compare the 25
candidates the cosine-similarity split designates high-fit against the
25 candidates that evaluator's own CV-only score would rank highest,
per job, and report the mean overlap (out of 25) across the ten jobs.

\begin{table}[h]
\centering
\caption{Agreement between the fixed cosine-similarity high-fit split
  and the high-fit split each evaluator's own independent CV-only score
  would draw instead. Overlap is the mean number (out of 25) of
  cosine-similarity-designated high-fit candidates that also fall in
  that evaluator's own top 25 by CV-only score, averaged across the ten
  jobs. Evaluators are ordered by decreasing agreement.}
\label{tab:cosine_agreement}
\begin{tabular}{@{}lcc@{}}
\toprule
Evaluator & Overlap (of 25) & Agreement \\
\midrule
Claude Sonnet 4.6  & 23.4 & 93.6\% \\
Claude Haiku 4.5   & 23.2 & 92.8\% \\
GPT-5              & 22.5 & 90.0\% \\
Gemini 3 Flash     & 22.5 & 90.0\% \\
GPT-4o mini        & 22.5 & 90.0\% \\
Gemini 3.5 Flash   & 22.4 & 89.6\% \\
DeepSeek V4 Flash  & 22.2 & 88.8\% \\
GPT-5 mini         & 21.9 & 87.6\% \\
DeepSeek V3.2      & 21.5 & 86.0\% \\
\midrule
\textbf{Mean}      & \textbf{22.5} & \textbf{90.0\%} \\
\bottomrule
\end{tabular}
\end{table}

The two Claude models agree most closely with the fixed split (93--94\%);
the two DeepSeek models agree least (86--89\%), with the OpenAI and
Gemini models clustered in between (88--90\%). No evaluator is a
dramatic outlier, and even the loosest match (DeepSeek~V3.2) still
agrees with the fixed split on 86\% of candidates — support for using
one shared criterion throughout the paper.

\subsection{Writer Preference Agreement Across Evaluators}\label{secA6}

\begin{figure}[!htbp]
  \centering
  \begin{minipage}[t]{0.48\textwidth}
    \centering
    \includegraphics[width=\textwidth]{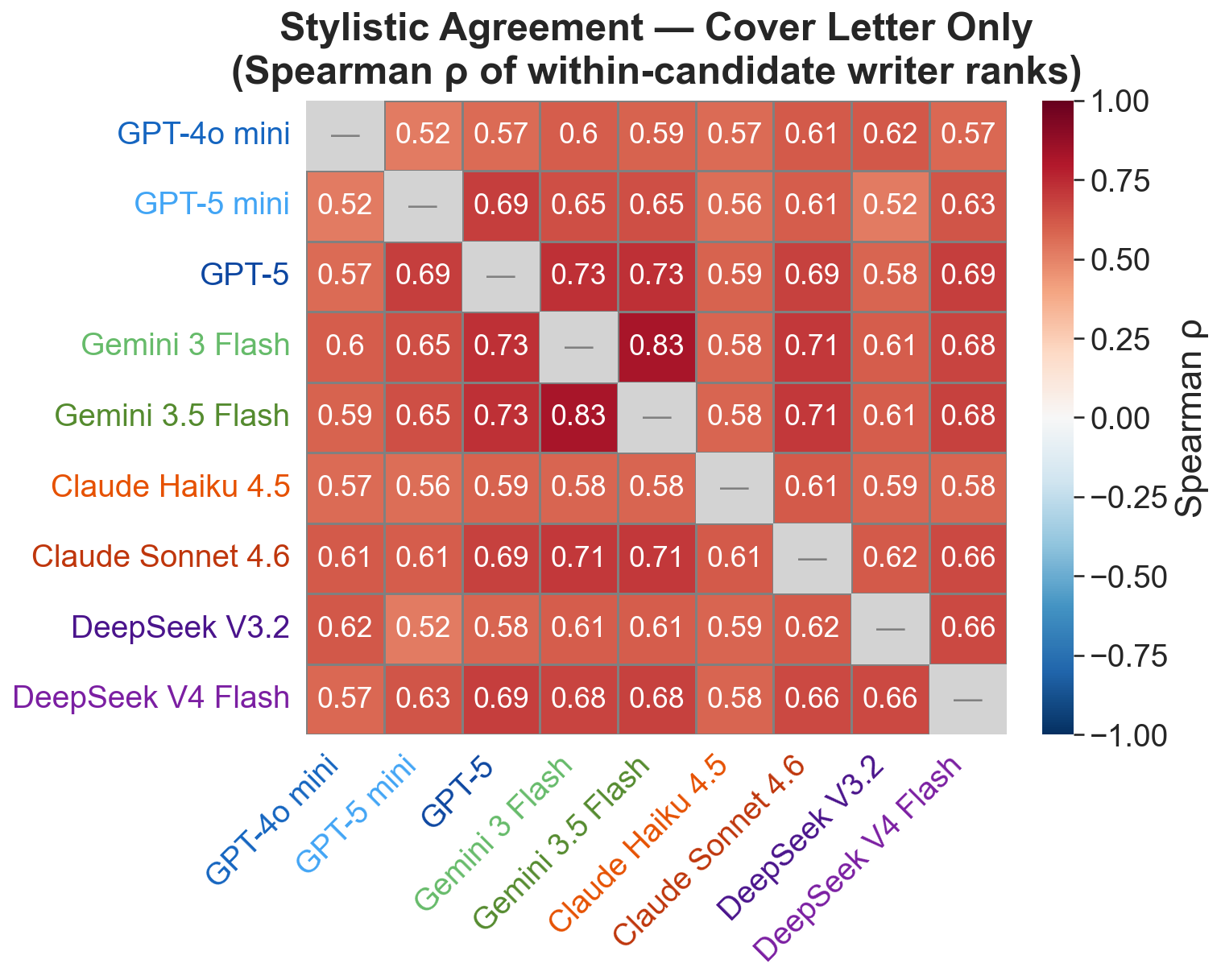}
  \end{minipage}
  \hfill
  \begin{minipage}[t]{0.48\textwidth}
    \centering
    \includegraphics[width=\textwidth]{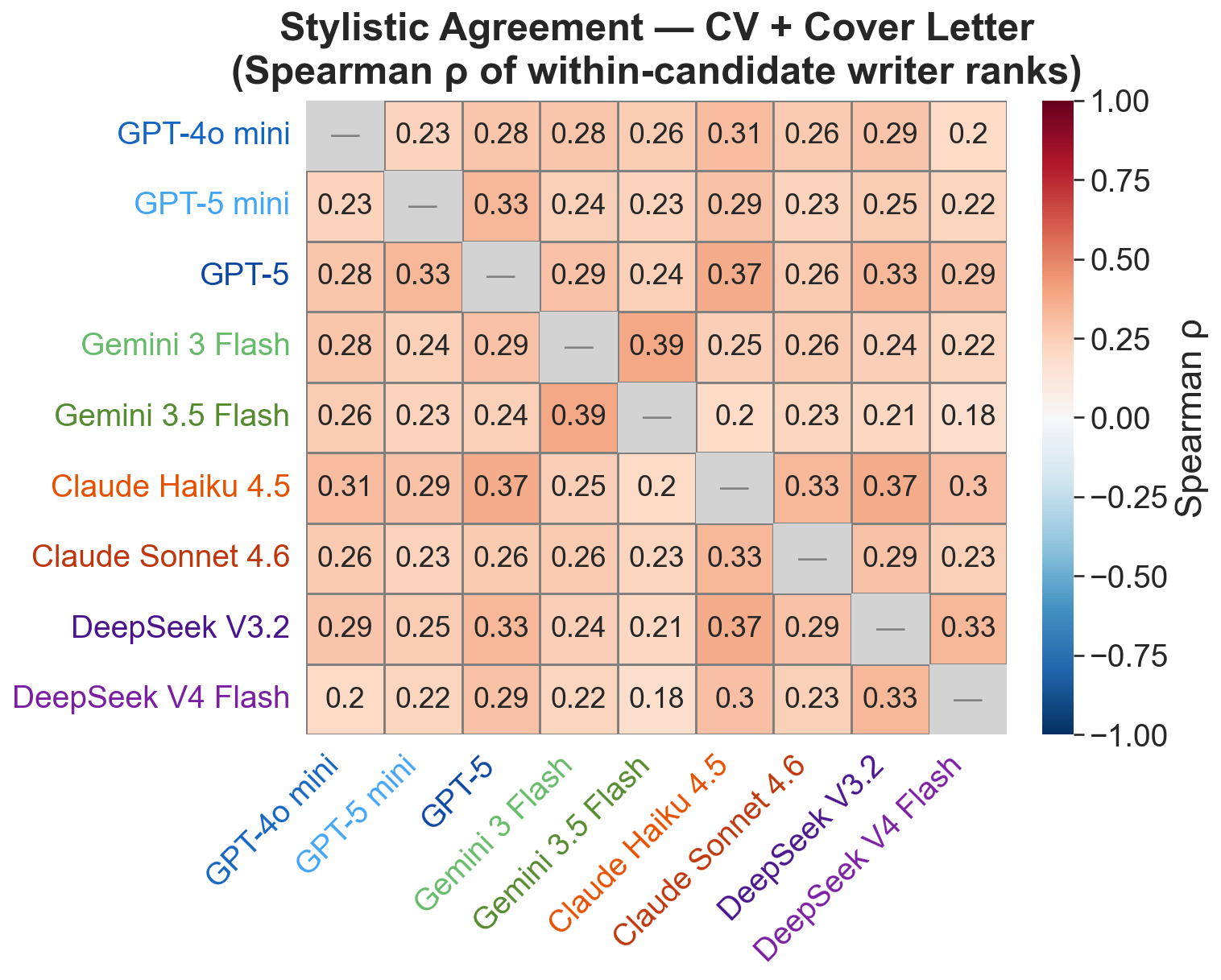}
  \end{minipage}
  \caption{Writer preference agreement: Spearman $\rho$ on within-(job,
    candidate) writer ranks $r_{e,j,c,w} = \mathrm{rank}_w(s_{e,j,c,w})$.
    By ranking within each evaluation context, this removes absolute
    strictness, candidate quality, and job difficulty simultaneously.
    \textbf{Left:} cover-letter-only condition (range 0.52--0.83, mean
    $\bar{\rho} = 0.63$). \textbf{Right:} combined CV + Cover letter
    condition (range 0.18--0.39, mean $\bar{\rho} = 0.27$).}
  \label{fig:agree_style}
\end{figure}

For each (job $j$, candidate $c$, evaluator $e$) triple we rank the
eleven writer models by score, from most to least preferred, giving a
length-11 ranking vector. We then compute the Spearman $\rho$ between
two evaluators' ranking vectors for the same (job, candidate) pair and
average across all pairs, which tells us how often two evaluators
agree on which writer produced the best letter for the same candidate.
Figure~\ref{fig:agree_style} shows the results (summarised in
Section~\ref{sec:preference_hierarchy}): agreement is well above chance
but far from perfect with the cover letter alone ($\bar\rho=0.63$), and
drops sharply once the CV is added ($\bar\rho=0.27$). That drop is not
because evaluators become less reliable: each evaluator's own
judgment of candidate quality is, if anything, just as consistent in
the combined condition ($\bar\rho=0.91$) as CV-only ($\bar\rho=0.92$;
both measured directly on scores, Figure~\ref{fig:strictness_by_condition}(b)),
but because the CV now supplies most of the discriminating signal,
leaving only a small, noisier residual driven by writer identity.

\section{Net Competitive Advantage by Individual Evaluator}\label{secB}

\begin{figure}[!htbp]
  \centering
  \includegraphics[width=\textwidth]{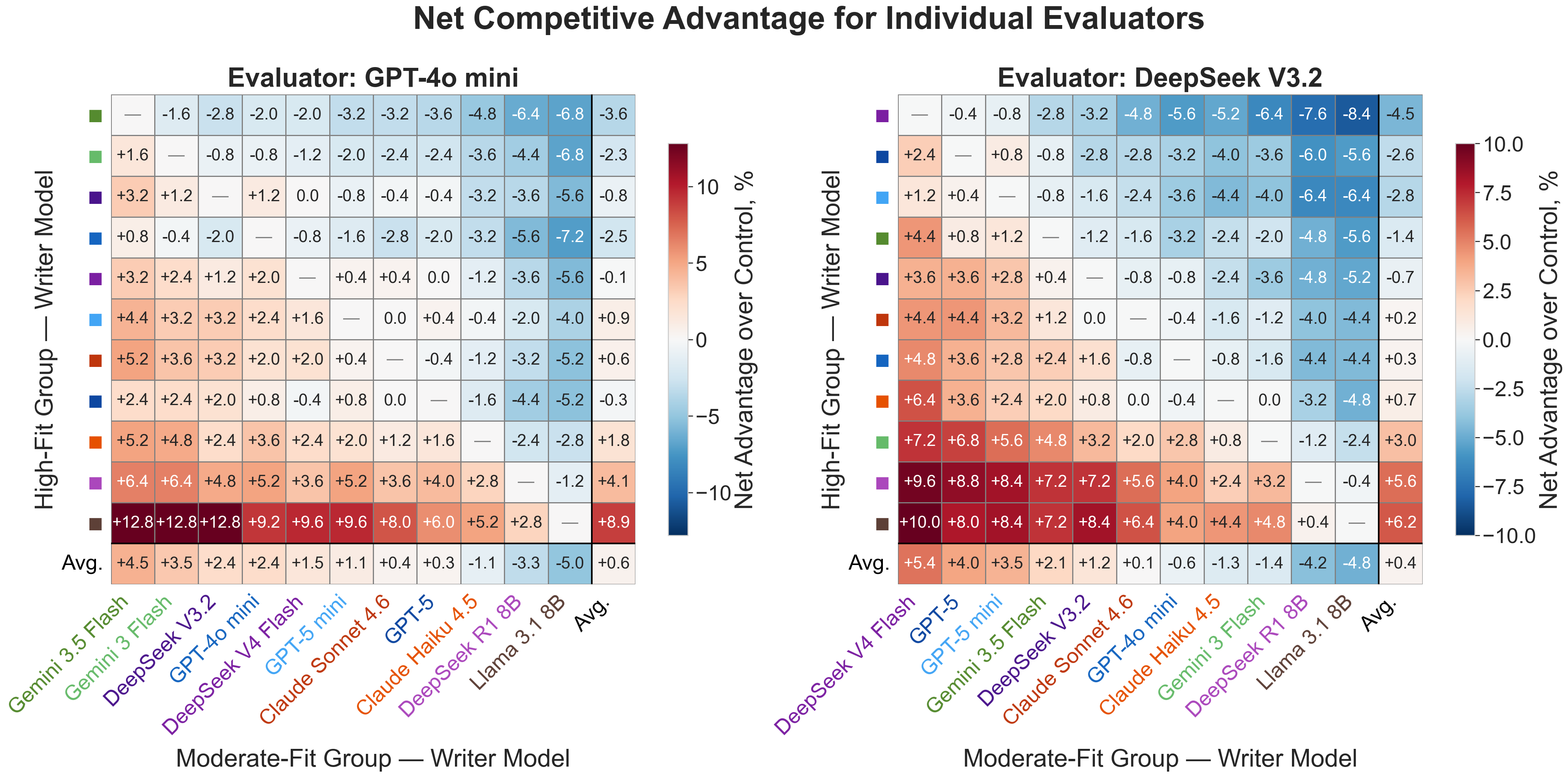}
  \caption{Net competitive advantage (as in Figure~\ref{fig:win_nca},
    right panel) for two individual evaluators rather than pooled
    across all nine: GPT-4o-mini (left), which has the single most
    extreme cell ($+12.8\%$) of any evaluator, and DeepSeek~V3.2
    (right), which has the highest total displacement summed across its
    whole matrix. Grey ``---'' marks the same-writer control diagonal.}
  \label{fig:net_extreme}
\end{figure}

Section~\ref{sec:competitiveness} reports the net displacement rate
pooled across all nine evaluators ($+7.3\%$ for the most adverse
pairing, Llama-3.1-8B vs.\ DeepSeek~V3.2, $\approx 1.8$ of 25 high-fit
candidates). Pooling averages over real evaluator-to-evaluator
variation: recomputing the same net displacement rate separately per
evaluator (self-evaluation triples included throughout, as in the
pooled calculation) gives a mean $\pm$ std of $+7.3\% \pm 3.2\%$ for
this pairing specifically, and individual evaluators can go well
beyond the pooled figure. Figure~\ref{fig:net_extreme} shows the two
most extreme individual evaluators: GPT-4o-mini, which reaches the
single highest net displacement cell across the whole grid ($+12.8\%$,
$\approx 3.2$ of 25 candidates, for three different challenger writers
against Llama-3.1-8B), and DeepSeek~V3.2, which has the highest total
displacement summed in absolute value across its whole matrix,
reflecting a broader pattern of elevated displacement rather than one
extreme pairing. Under either of these two evaluators specifically, the
worst-case harm to high-fit candidates reaches upwards of three per
job.

\section{Explaining Preference Through Writing-Style
Features}\label{secC}

This appendix gives the full detail behind
Section~\ref{sec:evaluator_explainability}: scatter-level and
per-writer/per-evaluator detail behind the feature-level regression of
Section~\ref{sec:explain_regression} (\ref{secC1}, \ref{secC2}), and
the variance-decomposition completeness check (\ref{secC3}).

\subsection{Writing Behavior vs. Evaluation Preference}\label{secC1}

\begin{figure}[!htbp]
  \centering
  \includegraphics[width=0.95\textwidth]{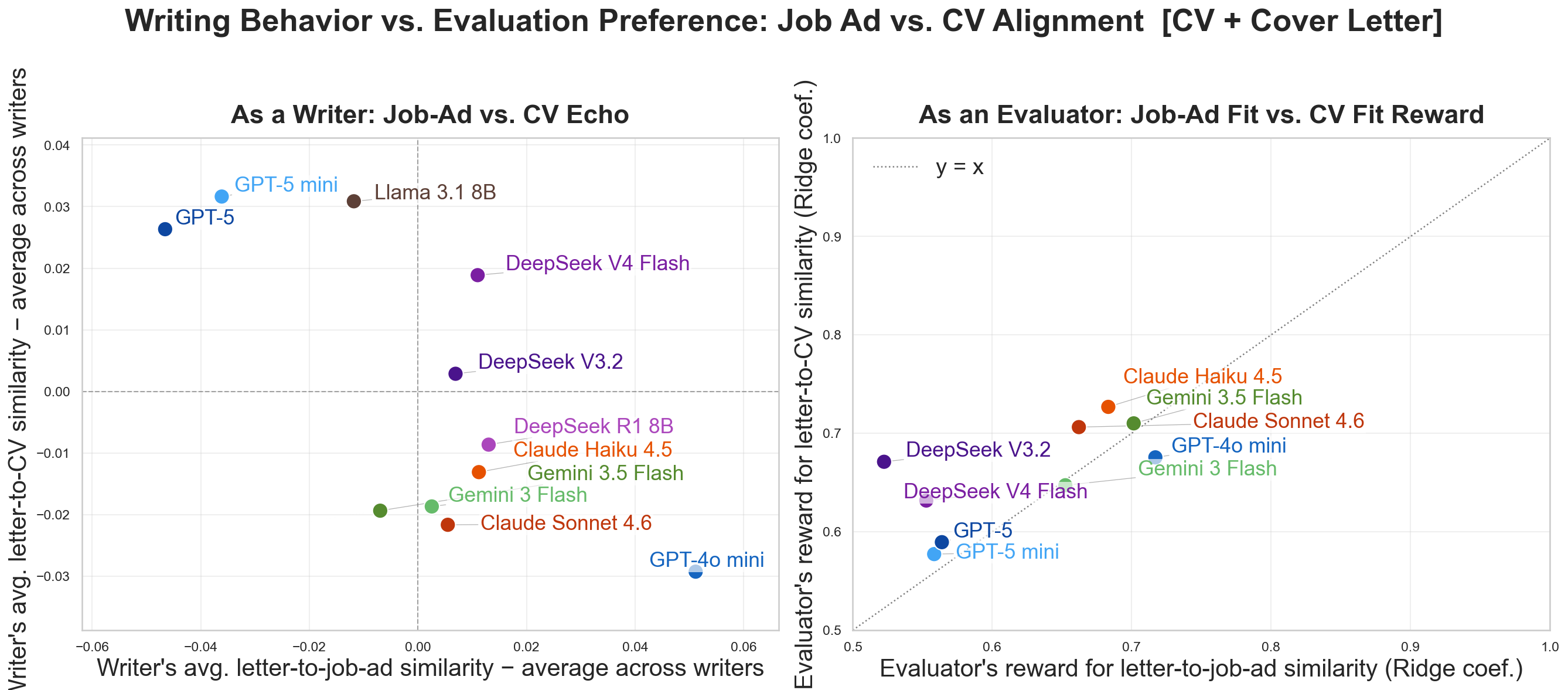}
  \caption{CV + Cover Letter condition. Left: each writer's average
    letter-to-job-ad vs.\ letter-to-CV cosine similarity, centred on the
    across-writer mean; GPT-4o-mini is the clearest job-ad-leaning
    outlier, GPT-5 and GPT-5-mini the clearest CV-leaning ones. Right:
    each evaluator's Ridge coefficient rewarding job-ad fit vs.\ CV
    fit. Points cluster tightly around the $y=x$ diagonal (fitted slope
    $0.49$), visually confirming that the two forms of semantic fit are
    rewarded almost identically once the CV is actually visible.}
  \label{fig:writer_vs_eval_cosine}
\end{figure}

\paragraph{Job ad vs.\ CV alignment, writer and evaluator side by side.}
Figure~\ref{fig:writer_vs_eval_cosine} pairs what writers actually do
(left) with what evaluators reward (right), for the two semantic-fit
features together. Shown for the CV~+~Cover Letter condition only: the
evaluator panel requires a coefficient on Cosine Similarity to CV, which, as
discussed in Section~\ref{sec:explain_regression}, is not a valid
regressor in the cover-letter-only condition.

\begin{figure}[!htbp]
  \centering
  \includegraphics[width=0.95\textwidth]{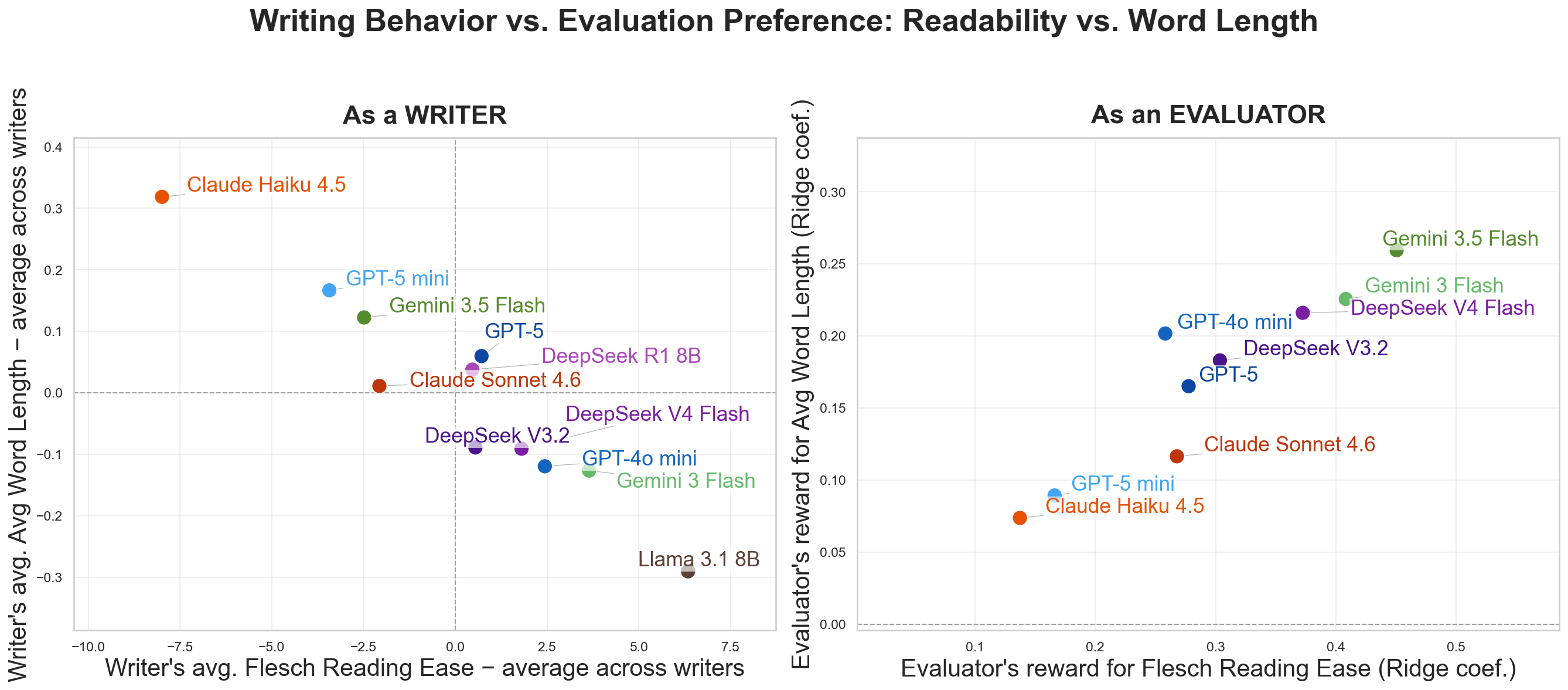}
  \caption{As writers, readability and average word length trade off
    sharply: no writer combines long, complex words with easy-to-read
    structure. As evaluators, that constraint disappears: every model
    rewards \emph{both} higher readability and longer words
    simultaneously, a combination essentially no writer among the eleven
    actually produces.}
  \label{fig:writer_vs_eval_flesch}
\end{figure}

\begin{figure}[!htbp]
  \centering
  \includegraphics[width=0.95\textwidth]{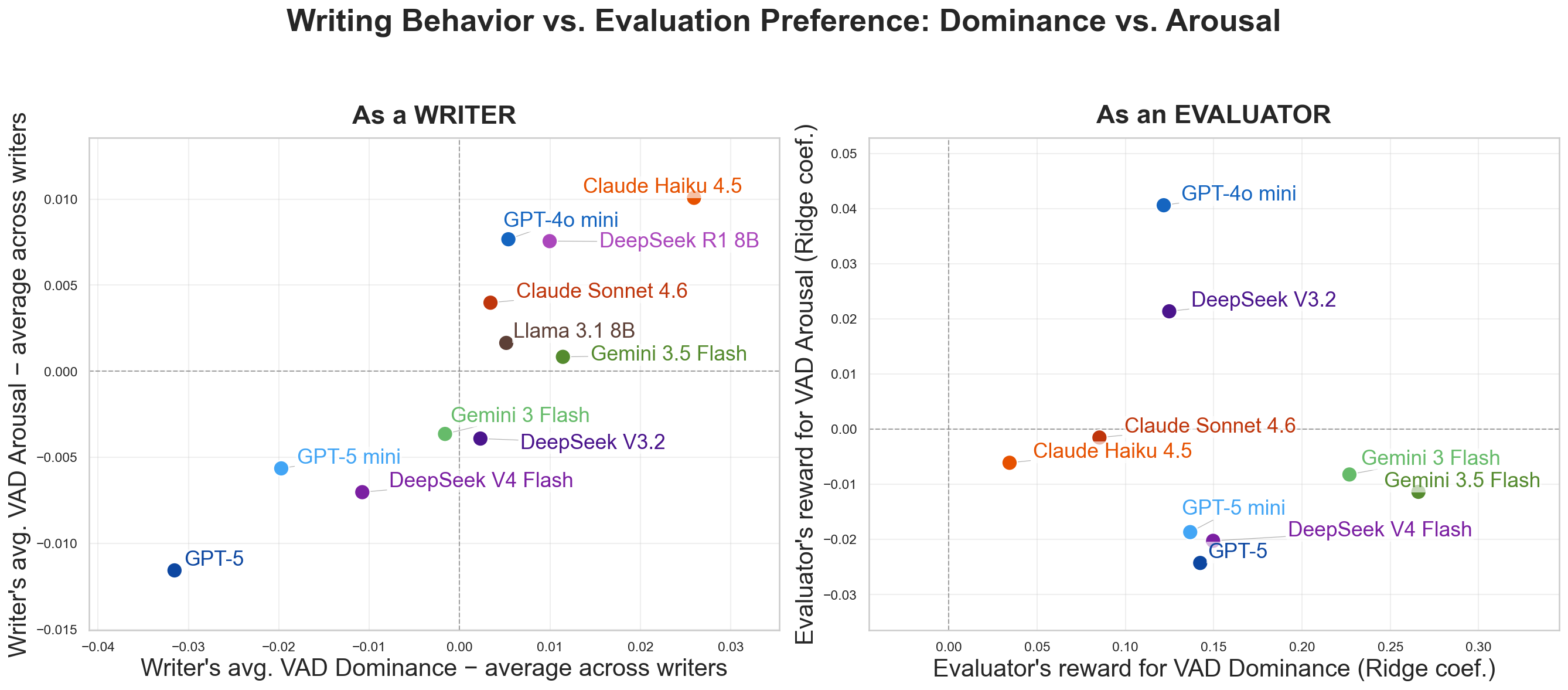}
  \caption{As writers, dominance and arousal move together: a more
    assertive-sounding letter also tends to read as more energetic. As
    evaluators, the relationship partially reverses: every model
    rewards dominance, but the ones that reward it most tend to reward
    arousal least, favouring "confident but composed" language over
    "assertive and energetic."}
  \label{fig:writer_vs_eval_dominance}
\end{figure}

\begin{figure}[!htbp]
  \centering
  \includegraphics[width=0.95\textwidth]{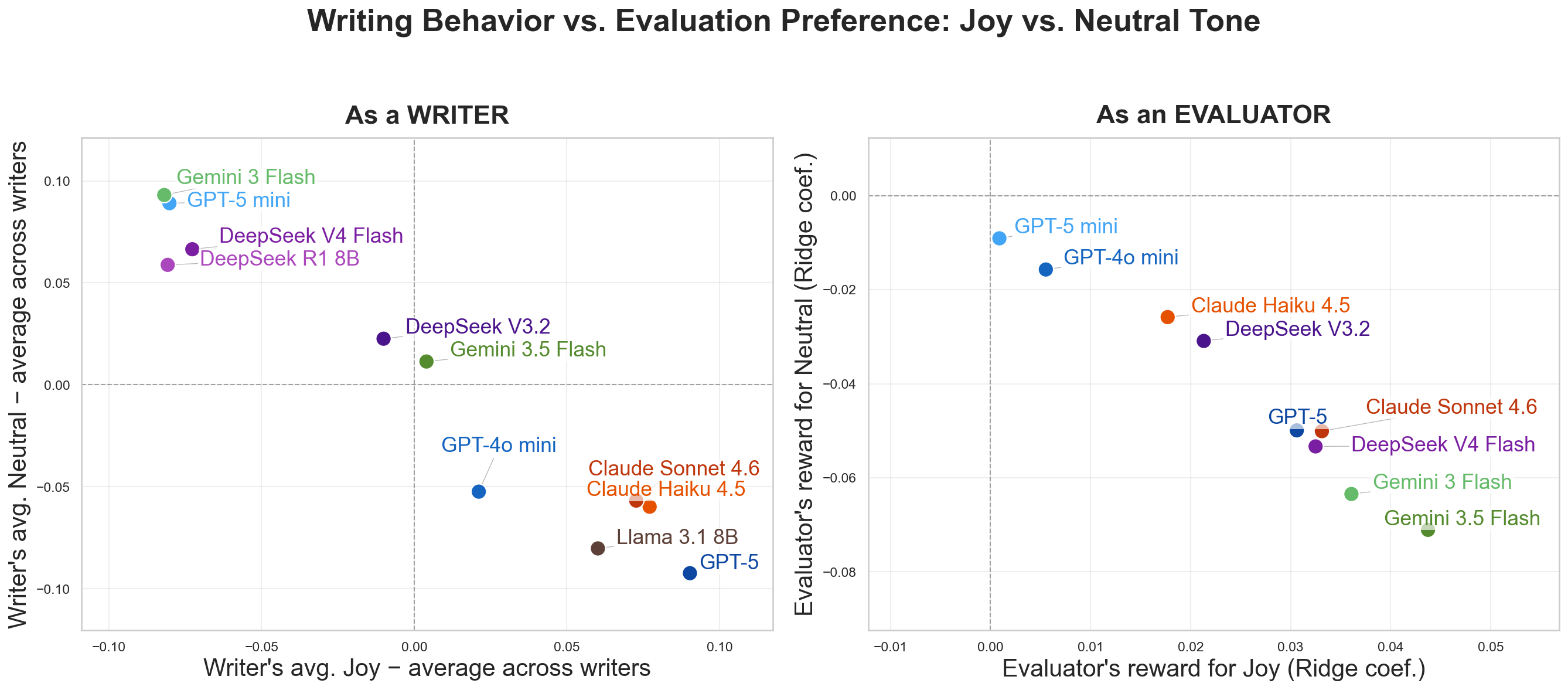}
  \caption{Both panels show a near-perfect trade-off between Joy and
    Neutral tone. The evaluator panel is not mechanically constrained to
    show this (Ridge coefficients on the two are fit independently), yet
    the mirror-image relationship is, if anything, steeper there than in
    how writers actually write: a genuine preference structure.}
  \label{fig:writer_vs_eval_joy}
\end{figure}

\paragraph{Trade-off pairs: what writers do vs.\ what evaluators want.}
Appendix~\ref{secA1} documents three within-writer style
trade-offs; Figures~\ref{fig:writer_vs_eval_flesch}--\ref{fig:writer_vs_eval_joy}
place evaluator preference alongside each one (cover-letter-only
condition; none of these features involve Cosine Similarity to CV).

\begin{figure}[!htbp]
  \centering
  \includegraphics[width=0.95\textwidth]{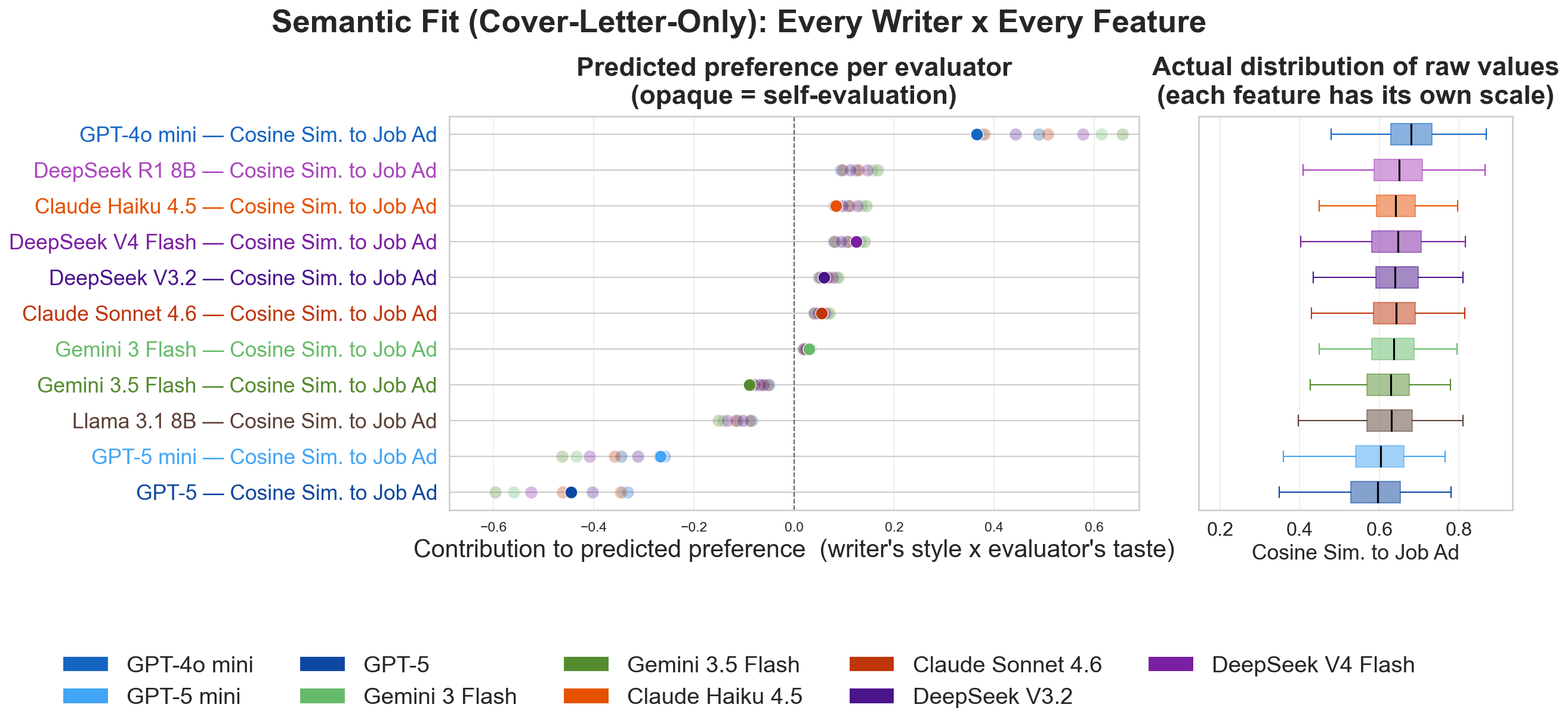}
  \caption{Cover-letter-only condition; only Cosine Similarity to Job Ad
    shown, since Cosine Similarity to CV is excluded (Section~\ref{sec:explain_regression}).
    GPT-4o-mini's own dot (opaque, top row) sits at the \emph{low} end
    of its own row's spread: every other evaluator rewards its job-ad
    echo more than GPT-4o-mini itself does.}
  \label{fig:semantic_fit_clonly}
\end{figure}

\begin{figure}[!htbp]
  \centering
  \includegraphics[width=0.95\textwidth]{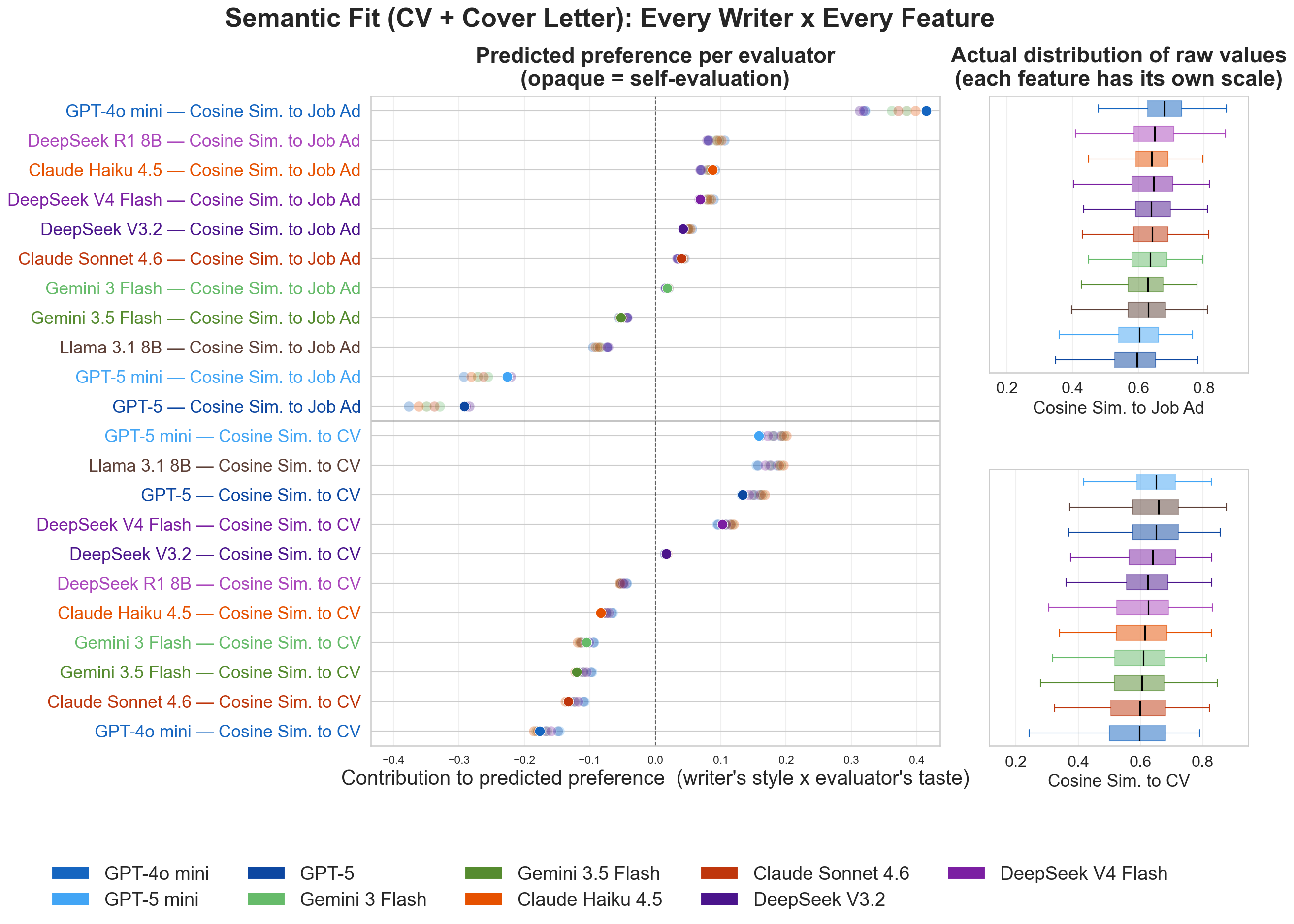}
  \caption{CV + Cover Letter condition; both semantic-fit features
    shown. Here GPT-4o-mini's own dot on Cosine Similarity to Job Ad sits at
    the \emph{high} end of its row instead: the self-preference
    documented in Section~\ref{sec:explain_regression} strengthens
    once the CV is present.}
  \label{fig:semantic_fit_cvcl}
\end{figure}

\paragraph{Semantic fit, every writer and every evaluator.}
Figures~\ref{fig:semantic_fit_clonly} and~\ref{fig:semantic_fit_cvcl}
give the full per-writer, per-evaluator breakdown behind the Cosine
Similarity to Job Ad (and, in the combined condition, Cosine Similarity to CV)
rows of Figure~\ref{fig:common_pref_ranked}, with every evaluator's dot
visible (opaque = self-evaluation).

\begin{figure}[!htbp]
  \centering
  \includegraphics[width=0.95\textwidth]{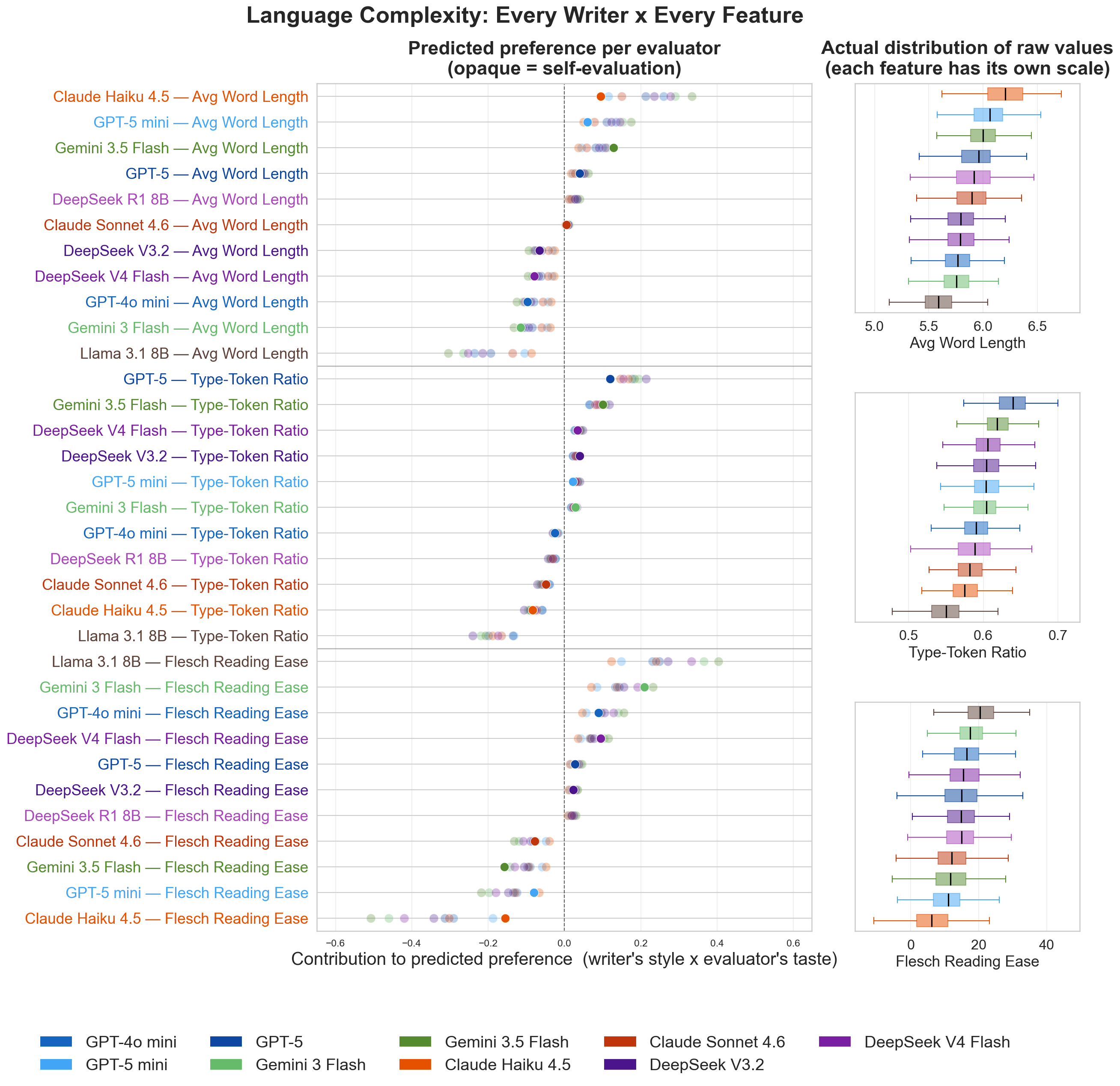}
  \caption{Cover-letter-only condition. Confirms Section~\ref{sec:explain_regression}'s
    reading at the individual-evaluator level: Claude~Haiku~4.5's long
    average word length and Llama-3.1-8B's high readability are each
    rewarded consistently across (almost) every evaluator.}
  \label{fig:language_complexity_cat}
\end{figure}

\paragraph{Language complexity, every writer and every evaluator.}
Figure~\ref{fig:language_complexity_cat} gives the same per-writer,
per-evaluator detail for Average Word Length, Type-Token Ratio, and Flesch
Reading Ease (cover-letter-only condition; none of these features
involve Cosine Similarity to CV).

\subsection{Feature Preference and Self-Preference Under CV + Cover Letter}\label{secC2}

\begin{figure}[!htbp]
  \centering
  \includegraphics[width=0.95\textwidth]{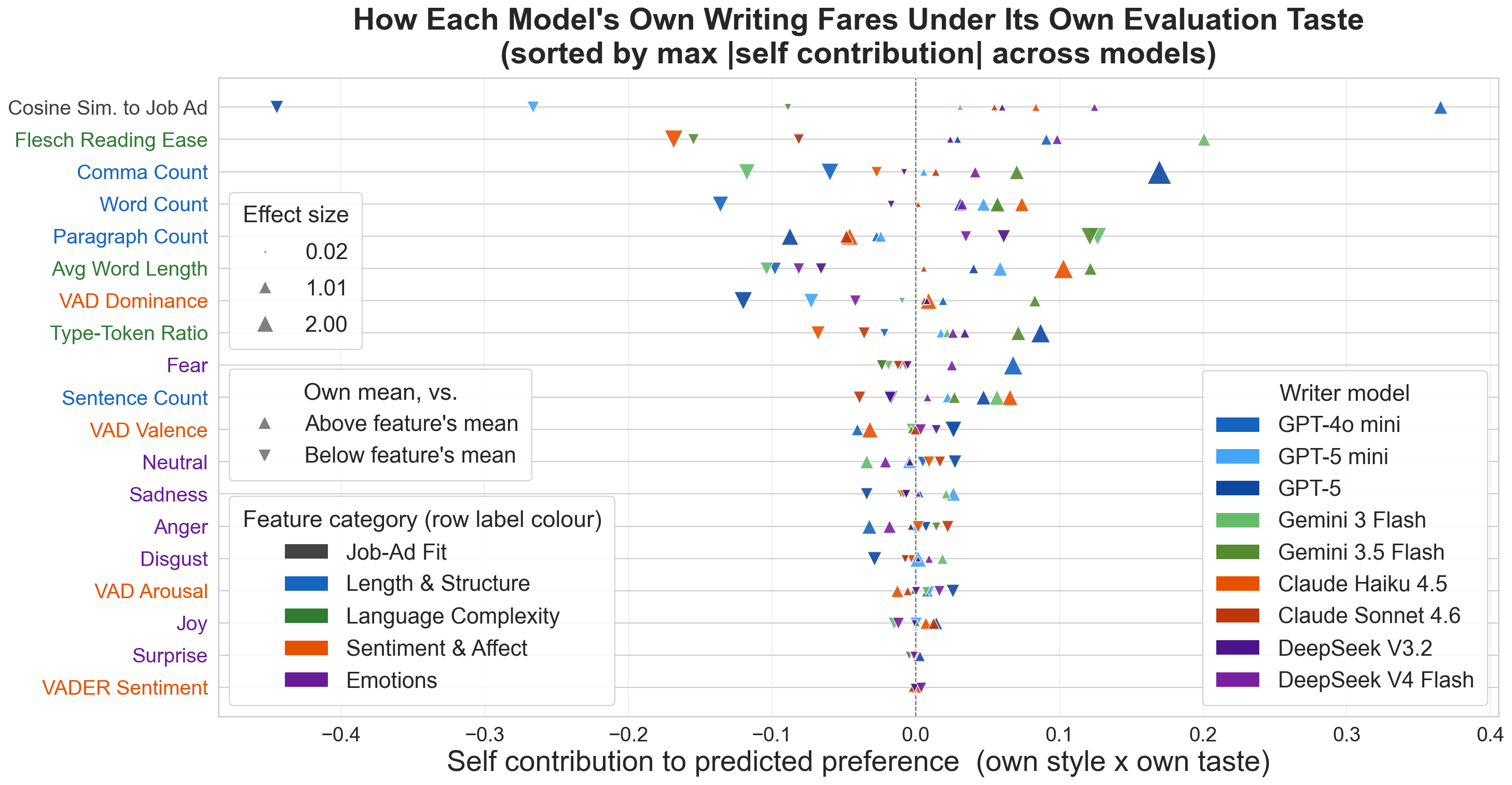}
  \caption{Self-evaluation diagonal only (cover-letter-only condition):
    for each feature, one triangle per dual-role model at its own
    contribution to its own predicted preference (own standardised
    style $\times$ own Ridge coefficient), ranked top to bottom by the
    largest $|\text{value}|$ across models. Triangle direction shows
    whether that model's own mean sits above ($\blacktriangle$) or
    below ($\blacktriangledown$) the feature's overall mean; size shows
    the standardised effect size of that deviation. Row labels are
    coloured by feature category as in Figure~\ref{fig:common_pref_ranked}.}
  \label{fig:self_points_ranked}
\end{figure}

Section~\ref{sec:explain_regression} reports Figure~\ref{fig:common_pref_ranked}
for the cover-letter-only condition; its self-preference companion,
Figure~\ref{fig:self_points_ranked}, and the full self-preference
analysis behind it appear here, alongside both figures' CV~+~Cover
Letter counterparts, to check whether the main text's conclusions are
an artifact of withholding the CV from the evaluator.

\paragraph{Feature-level preference.}
The overall picture holds: semantic fit still dominates, and the same
broad set of length and complexity features remain unanimously
rewarded, with Paragraph Count and Anger still among the unanimous
penalties. The clearest difference is the reappearance of Cosine
Sim.\ to CV itself: excluded from the cover-letter-only regression
entirely (Section~\ref{sec:explain_regression}), it re-enters once
the evaluator can actually see the CV and immediately becomes, by a
wide margin, one of the two most rewarded features, essentially tied
with Cosine Similarity to Job Ad ($0.60$ vs.\ $0.60$), both well separated
from every style feature. Plotted against each other
(Figure~\ref{fig:writer_vs_eval_cosine}, Appendix~\ref{secC1}),
evaluators' reward for job-ad fit and CV fit cluster tightly around the
$y=x$ diagonal in this condition. In the cover-letter-only case, by
contrast, the points sit uniformly on one side of it. Evaluator consensus also
shifts on several affect features: VAD Dominance, unanimously rewarded
cover-letter-only ($+0.11$), loses that status once the CV is present
(mean drops to $+0.05$, one evaluator now penalises it), while VADER
Sentiment flips from no consensus to a small unanimous penalty
($-0.05$); VAD Arousal remains without consensus in both conditions. A
plausible reading is that once genuine candidate information is
available, evaluators lean even more heavily on the two directly
verifiable semantic-fit signals, leaving affect-based style cues less
room to move the score consistently in one direction.

\paragraph{Self-preference (cover-letter-only).}
Figure~\ref{fig:self_points_ranked} applies the same standardised
effect-size lens specifically to self-evaluation: for each feature,
only the evaluator-equals-writer diagonal is shown, ranked by the
largest $|\text{self contribution}|$ across the nine dual-role models.
The single most extreme self-preference and self-penalty in the entire
figure both occur on the feature identified above as the most
universally rewarded: GPT-4o-mini's self-contribution from Cosine
Similarity to Job Ad is $+0.36$, while GPT-5's is $-0.45$: the two
writers whose own usage of that feature is most extreme
(Figure~\ref{fig:common_pref_ranked}, right panel) are also,
unsurprisingly, the two most affected by it under their own evaluation
taste, in opposite directions. GPT-4o-mini's case is a product, not a
show of generosity: its own Ridge coefficient for this feature
($+0.60$) is in fact the \emph{lowest} of all nine evaluators
(Section~\ref{sec:explain_regression}), but its usage is so extreme
that usage times even a comparatively modest self-reward still yields
the largest self-preference contribution in the figure. Self-preference is not confined to
semantic fit, however: on Flesch Reading Ease, Gemini~3~Flash rewards
its own high readability most ($+0.20$), while Claude~Haiku~4.5 and
Gemini~3.5~Flash are both strongly self-critical about their own low
readability ($-0.17$ and $-0.15$ respectively). The same mechanism can
cut against a writer that otherwise benefits from it elsewhere:
GPT-4o-mini, whose semantic-fit self-preference is the largest in the
entire figure, is simultaneously self-critical on Word Count ($-0.14$,
the most negative of any writer on that feature), because its own
letters are comparatively short on a feature every evaluator, including
itself, rewards. Self-preference, in other words, is feature-specific:
the same model can be the most
self-rewarding writer on one feature and among the most self-critical
on another.

\begin{figure}[!htbp]
  \centering
  \includegraphics[width=0.95\textwidth]{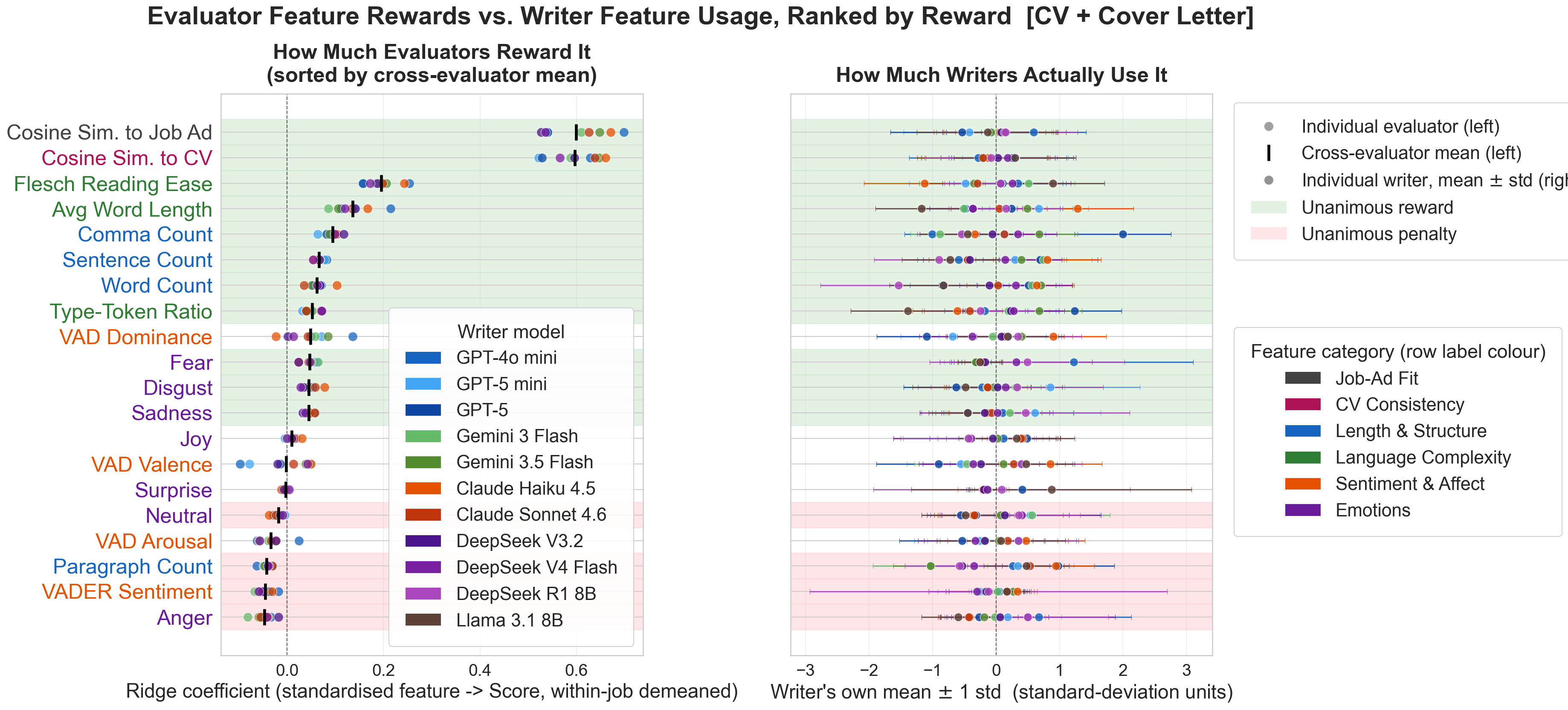}
  \caption{Same analysis as Figure~\ref{fig:common_pref_ranked}, for the
    CV + Cover Letter condition. Cosine Similarity to Job Ad and Cosine
    Sim.\ to CV converge to almost identical, dominant reward levels;
    VAD Dominance loses unanimous-reward status and VADER Sentiment
    gains unanimous-penalty status relative to the cover-letter-only
    condition.}
  \label{fig:common_pref_ranked_cvcl}
\end{figure}

\begin{figure}[!htbp]
  \centering
  \includegraphics[width=0.95\textwidth]{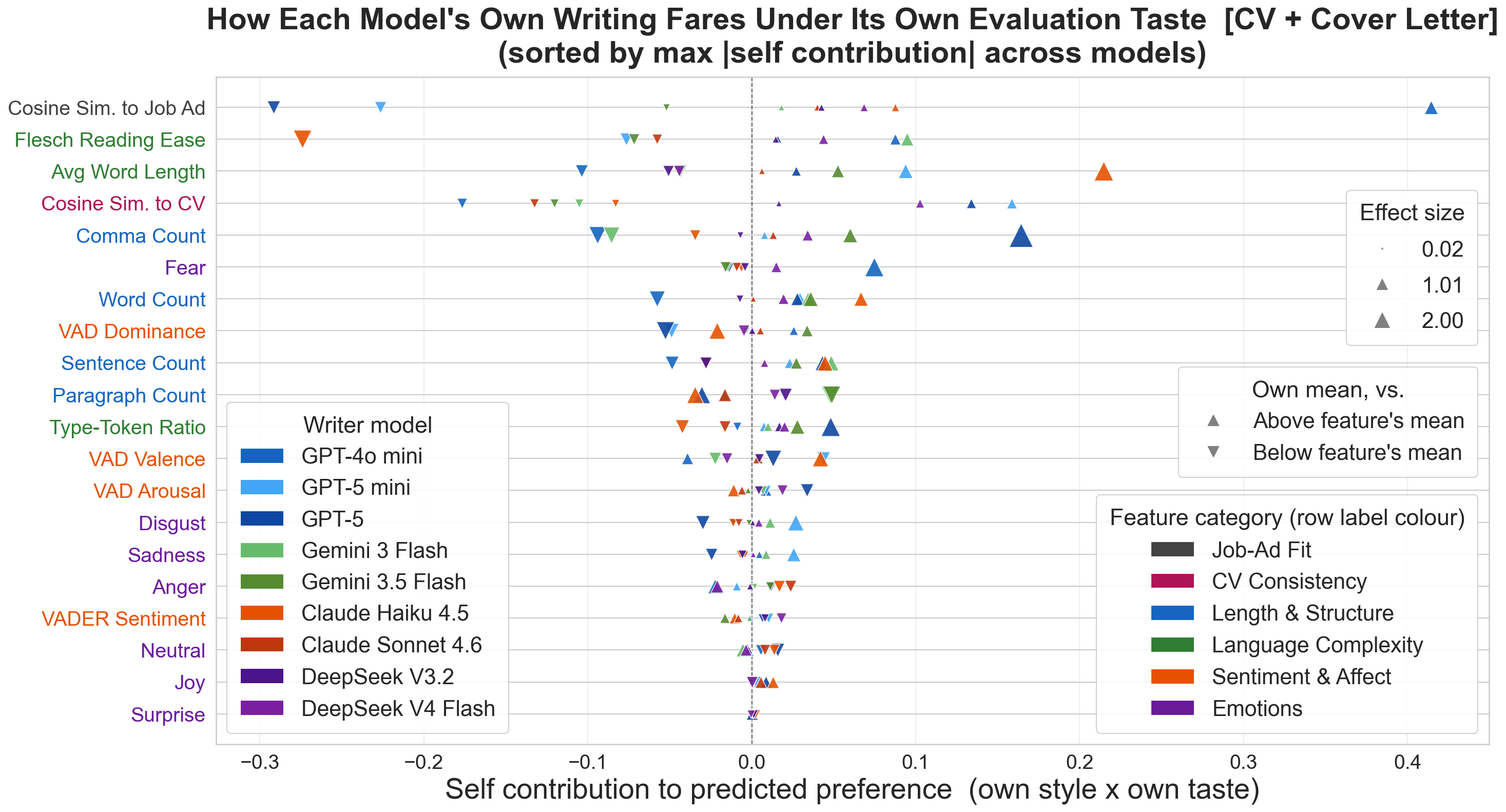}
  \caption{Same analysis as Figure~\ref{fig:self_points_ranked}, for the
    CV + Cover Letter condition. GPT-4o-mini's self-preference on
    Cosine Similarity to Job Ad grows further despite the feature's
    shrinking overall reward; Claude~Haiku~4.5's self-preference on
    Average Word Length emerges as the largest style-feature self-contribution
    in this condition.}
  \label{fig:self_points_ranked_cvcl}
\end{figure}

\paragraph{Self-preference under CV + Cover Letter.}
Figure~\ref{fig:self_points_ranked_cvcl} applies the same lens to the
CV~+~Cover Letter condition. The self-preference pattern is, if anything,
sharper. GPT-4o-mini's
self-preference on Cosine Similarity to Job Ad grows from $+0.36$
(cover-letter-only) to $+0.41$ (combined) even though the feature's
overall reward shrinks (Figure~\ref{fig:common_pref_ranked_cvcl}):
its own usage and its own taste for that feature both remain unusually
well aligned. The per-writer, per-evaluator detail behind this
(Figure~\ref{fig:semantic_fit_cvcl}, Appendix~\ref{secC1}) shows why:
in the cover-letter-only version
(Figure~\ref{fig:semantic_fit_clonly}, Appendix~\ref{secC1}),
GPT-4o-mini's own dot on Cosine Similarity to Job Ad sat at the low end of
its row; now it sits at the \emph{high} end. GPT-5 remains the
most self-critical on the same feature ($-0.29$, vs.\ $-0.45$
cover-letter-only). The average-word-length / readability trade-off
documented in Appendix~\ref{secA1} shows up here too:
Claude~Haiku~4.5's self-preference on Average Word Length grows to
$+0.22$, the largest style-feature self-contribution in the
combined-condition figure, reflecting the same longest-words-of-any-writer
trait.

\begin{figure}[!htbp]
  \centering
  \includegraphics[width=0.98\textwidth]{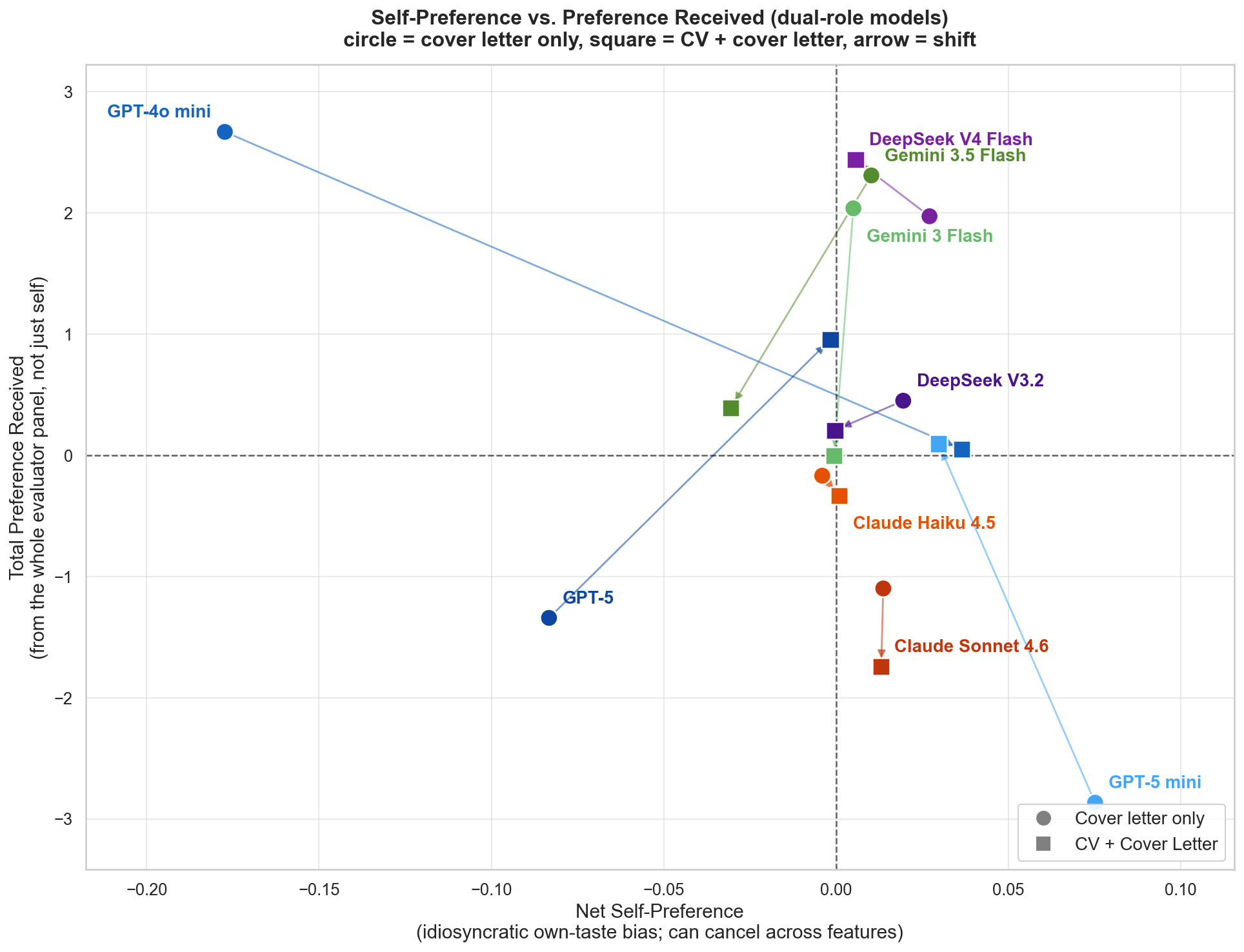}
  \caption{Cover-letter-only (circle) and CV~+~Cover Letter (square)
    conditions, Cosine Similarity to CV excluded from the cover-letter-only
    fit (see text). One point per condition per dual-role model (the
    nine that are both writer and evaluator; Net Self-Preference is
    undefined for the two locally-hosted writers, which are never
    evaluators and so do not appear here), connected by an arrow from
    cover-letter-only to CV~+~Cover~Letter: $x$-axis, Net
    Self-Preference, the signed sum of the model's own idiosyncratic
    bias across all measured features; $y$-axis, Total Preference
    Received, that writer's own style dotted with the summed Ridge
    coefficients of all nine evaluators. In the cover-letter-only
    condition, GPT-4o-mini and GPT-5-mini sit at opposite extremes:
    GPT-4o-mini is the harshest critic of its own writing yet the most
    liked by the panel overall, while GPT-5-mini is the mirror image.
    Both collapse sharply toward the origin once the CV is present:
    GPT-4o-mini's Net Self-Preference turns slightly positive
    ($+0.036$) while its Total Preference Received falls to near zero
    ($+0.05$, down from $+2.98$), and DeepSeek~V4~Flash becomes the
    writer most liked by the panel overall ($+2.44$).}
  \label{fig:self_pref_received}
\end{figure}

\paragraph{Net self-preference vs.\ total preference received.}
Figure~\ref{fig:self_pref_received} plots these two quantities against
each other for every dual-role model. Section~\ref{sec:explain_regression}
summarizes each writer's overall
standing two ways: \emph{Net Self-Preference}, the signed sum of a
model's own idiosyncratic bias (own taste $-$ average taste, dotted
with its own style) across all measured features (nineteen
cover-letter-only, since Cosine Similarity to CV is not yet a valid
regressor; twenty once the CV is present), against \emph{Total
Preference Received}, the same writer's actual style dotted with the
summed taste of all nine evaluators, self included. Total Preference
Received is the feature-regression counterpart of the Relative
Preference Gap $\Delta_{\mathrm{pref}}(e,w)$ introduced in
Section~\ref{sec:preference_hierarchy} (Figure~\ref{fig:heatmap}):
where $\Delta_{\mathrm{pref}}$ is read directly off the raw scores,
Total Preference Received reconstructs the same
whole-panel-likes-this-writer quantity from the fitted per-feature
weights, and the two agree qualitatively (e.g.\ GPT-4o-mini tops both
in the cover-letter-only condition), which is itself a check that the
feature-level model is not missing whatever drives the raw preference
signal. Llama-3.1-8B, which is never an evaluator and so has no Net
Self-Preference score, finishes last on Total Preference Received
($-4.01$), consistent with its weak win rate
(Section~\ref{sec:competitiveness}) and Relative Preference Gap
(Section~\ref{sec:preference_hierarchy}).

\subsection{Variance Decomposition}\label{secC3}\label{sec:explain_variance}

\begin{figure}[!htbp]
  \centering
  \includegraphics[width=0.95\textwidth]{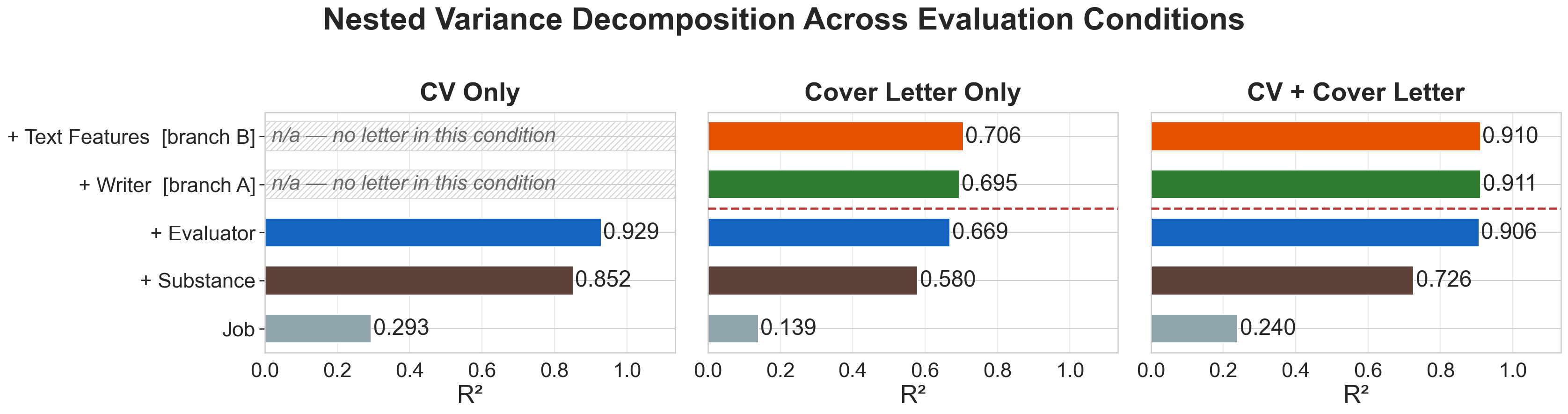}
  \caption{Nested variance decomposition, pooled across all nine
    evaluators, for CV Only, CV~+~Cover~Letter, and Cover-Letter-Only.
    Job~$\to$~Substance~$\to$~Evaluator is a shared trunk; the dashed
    red line marks where branch A (+~Writer) and branch B (+~Text
    Features) split off from ``+~Evaluator'' as two alternative,
    parallel explanations of the same remaining variance.
    Cover-Letter-Only's Text Features
    exclude letter-to-CV similarity, since the evaluator never sees
    the CV in that condition. CV Only has no letter, so branches A/B
    do not apply there (hatched rows). All panels share one $R^2$
    scale and row order.}
  \label{fig:nested_decomp}
\end{figure}

Section~\ref{sec:explain_regression} explained preference through
specific features; here we test whether that same feature set is
sufficient on its own, or whether writer identity carries additional
information once job, candidate substance, and each evaluator's own
baseline are accounted for. We use an exact Shapley (LMG) $R^2$
decomposition: each block of predictors' marginal contribution to
$R^2$ is averaged over every possible order in which the blocks could
be added to the model, so the resulting block-level shares sum exactly
to the total explained variance regardless of how correlated the
underlying blocks are. Figure~\ref{fig:nested_decomp} builds a shared
trunk (Job~$\to$~Substance~$\to$~Evaluator), pooled across all nine
evaluators, then branches into two alternative explanations of
whatever variance remains: adding Writer identity as a catch-all
categorical (branch A), or adding measured Text Features instead
(branch B). The trunk alone already reaches $R^2 = 0.906$ in
CV~+~Cover~Letter and $0.669$ in Cover-Letter-Only, so only a small
residual is left for the two branches to split. In CV~+~Cover~Letter
they split it almost evenly ($0.911$ vs.\ $0.910$, a $+0.004$
increment each): essentially everything Writer identity could explain
in that residual is already captured by measurable text properties. In
Cover-Letter-Only the split is less even: Text Features add more
than Writer identity does ($0.706$ vs.\ $0.695$, $+0.037$ vs.\
$+0.026$), but once the CV is invisible, whatever evaluators are
reacting to in ``who wrote this'' is, if anything, \emph{even more}
fully captured by measurable writing properties than by raw identity.
(CV~Only has no letter, so branches A/B do not apply there; its trunk
alone already reaches $R^2 = 0.929$, unsurprising since the outcome
being predicted \emph{is} essentially that same holistic judgment.)
Either way, there is little evidence of an unmeasured ``writer
fingerprint'' beyond what the feature set already captures, though the
residual being split is itself small relative to what Job, Substance,
and Evaluator identity already explain.

We also break this decomposition down per evaluator, in four ways:
the per-evaluator Substance/Writer split, the per-evaluator
Substance/style split, the robustness checks behind the
raw-score-to-rank explanation of the Writer-variance/rank-gap-bias
divergence, and the per-feature drilldown of style-driven variance.

\begin{figure}[!htbp]
  \centering
  \includegraphics[width=0.95\textwidth]{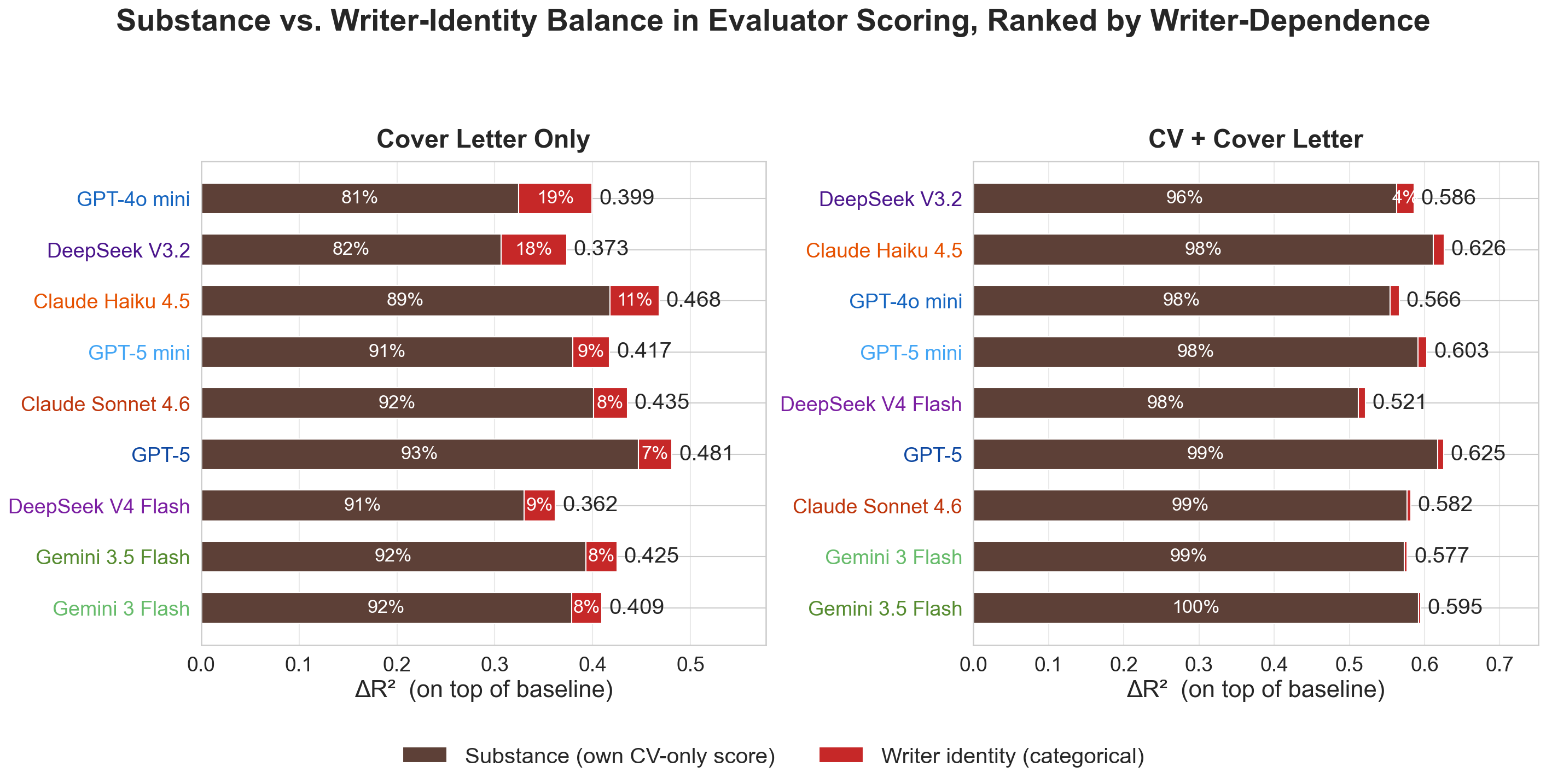}
  \caption{Exact Shapley split of $R^2$ on top of a job-only baseline,
    per evaluator, using only two blocks, Substance (that
    evaluator's own CV-only score) and Writer identity (categorical),
    and ranked by each evaluator's absolute Writer contribution. A
    large Writer share relative to Substance is a direct,
    regression-based measure of how little of an evaluator's score
    tracks genuine candidate quality once its own substance judgment
    is already accounted for.}
  \label{fig:substance_writer}
\end{figure}

\paragraph{Substance vs.\ Writer, per evaluator.}
Figure~\ref{fig:substance_writer} isolates the question
Section~\ref{sec:evaluator_fairness}'s rank-based metric answered only
indirectly: with each evaluator's own Substance judgment as the sole
genuine-quality control (no semantic or style blocks at all), how much
of the remaining explainable variance is attributable to raw Writer
identity? This is a direct, regression-based measure of how much of an
evaluator's \emph{own} score variance moves with writer identity,
ranked by each evaluator's absolute Writer contribution, and it
diverges from Figure~\ref{fig:fairness}'s rank-based bias ranking in
Cover-Letter-Only: GPT-4o~mini and DeepSeek~V3.2 have by far the
largest Writer contributions ($18.8\%$ and $17.9\%$ of their
explainable variance, respectively, more than any other evaluator by a
wide margin), yet both sit only mid-pack on the rank-based bias ranking
($10.51$ and $10.73$, ranks 4 and 5 of 9). Conversely, Gemini~3.5~Flash,
the \emph{most} rank-biased evaluator in this condition ($11.24$), has
one of the \emph{lowest} Writer variance shares ($7.5\%$). Once the CV
is also available, by contrast, the two lenses agree closely:
DeepSeek~V3.2's Writer block ($\Delta R^2 = 0.023$, $4\%$ of its total
explainable variance) is more than $1.6\times$ the next-highest
evaluator (Claude Haiku~4.5, $0.014$), matching its position as the
most biased evaluator by cumulative rank gap ($14.85$); Gemini~3.5~Flash
sits at the opposite extreme on both measures ($\Delta R^2 = 0.003$,
$0.4\%$; rank-gap $6.33$, the most neutral evaluator by either metric).
This condition is where the two metrics should agree most closely in
the first place: with the CV also present, the underlying $R^2$ is
high enough ($0.906$) that raw scores and their rank transform track
each other reliably, so the raw-score-to-rank distortion discussed
below has little room to operate. Figure~\ref{fig:substance_writer}
should be read as a diagnostic of each evaluator's own scoring
psychology (how much of \emph{its own} variability is writer-driven,
relative to everything else it reacts to), and not as a substitute
for, or a cross-check on, the cross-evaluator fairness ranking of
Section~\ref{sec:evaluator_fairness}.

\begin{figure}[!htbp]
  \centering
  \includegraphics[width=0.95\textwidth]{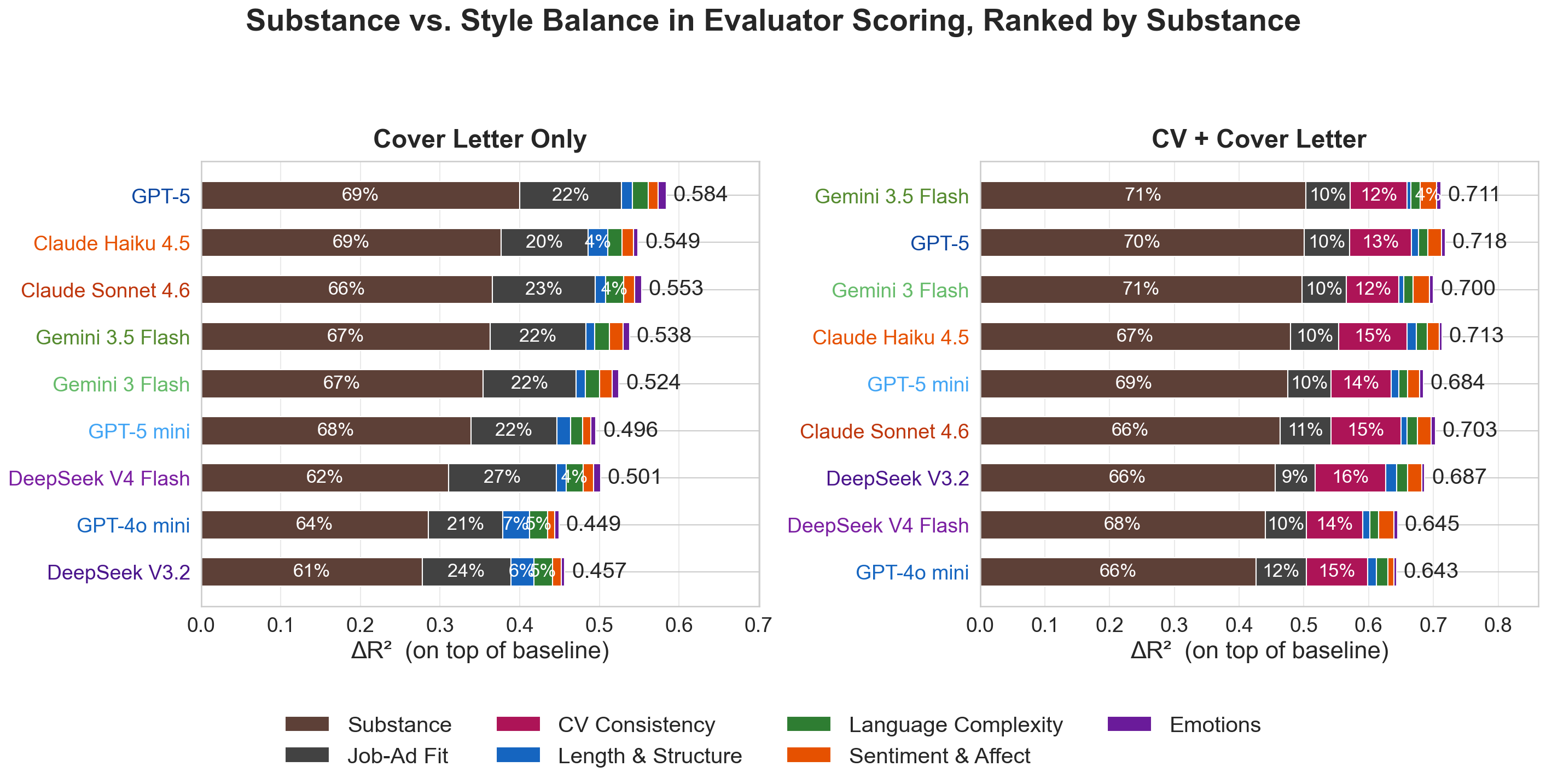}
  \caption{Exact Shapley split of $R^2$ on top of a job-only baseline,
    per evaluator, ranked by each evaluator's own Substance
    contribution. Substance is that evaluator's own CV-only score for
    the exact candidate being judged: a self-referential quantity
    specific to that evaluator and that candidate; since Writer identity is not a baseline
    control, between-writer variation is absorbed into the
    Semantic/style blocks.}
  \label{fig:substance_style}
\end{figure}

\paragraph{Substance vs.\ style, per evaluator.}
Figure~\ref{fig:substance_style} repeats the block-level Shapley
decomposition on a job-only baseline (Writer identity is deliberately
\emph{not} a baseline control here, so any between-writer variation
counts toward Semantic/style), ranked by each evaluator's own
Substance contribution alone, independent of total explained variance,
so the
ordering itself shows which evaluators anchor most heavily on their
own holistic candidate judgment. In Cover-Letter-Only, substance
shares range from $69\%$ (GPT-5, the most substance-anchored evaluator
in this condition) down to $61\%$ (DeepSeek~V3.2) and $64\%$
(GPT-4o~mini), the same two evaluators that lean most heavily on
writer identity once genuine candidate information is unavailable
(see below). Once the CV is available, the spread
narrows substantially and all nine evaluators become
substance-dominated and fairly close together, with
Gemini~3.5~Flash, GPT-5, and Gemini~3~Flash anchoring most heavily on
substance and GPT-4o~mini, Claude~Sonnet~4.6, and DeepSeek~V3.2
slightly less so.

\paragraph{Why the Writer-variance and rank-gap-bias metrics diverge
in Cover-Letter-Only.}
One might read the gap noted above skeptically, as a
measurement artifact: perhaps the $\Delta R^2$ regression's job-only
baseline is misspecified for Cover-Letter-Only in a way that inflates
or deflates the Writer share unevenly across evaluators, and a
better-controlled regression would bring the two metrics back into
line. We checked the two most obvious ways that could happen, and
neither holds up. Conditioning further on the exact candidate, beyond
just the job, rules out ``Gemini simply reacts more to genuine
candidate quality, diluting its Writer share'': it makes the
correlation with rank-gap bias \emph{worse}. Swapping the
normalized $\Delta R^2$ share for an equivalent \emph{unnormalized}
writer effect in raw-score units, which removes any denominator-driven
scale confound entirely, still only weakly tracks cumulative rank-gap
bias in Cover-Letter-Only (Spearman $\rho \approx 0.43$, not
significant, versus $\rho \approx 0.87$, $p = 0.002$, in
CV~+~Cover~Letter). What does explain it: recomputing that same
unnormalized quantity in \emph{rank} space instead of raw-score space
reproduces the rank-gap ranking almost exactly ($\rho \approx 1.0$).
So the gap traces back to the raw-score-to-rank transform itself.

\begin{figure}[!htbp]
  \centering
  \includegraphics[width=0.95\textwidth]{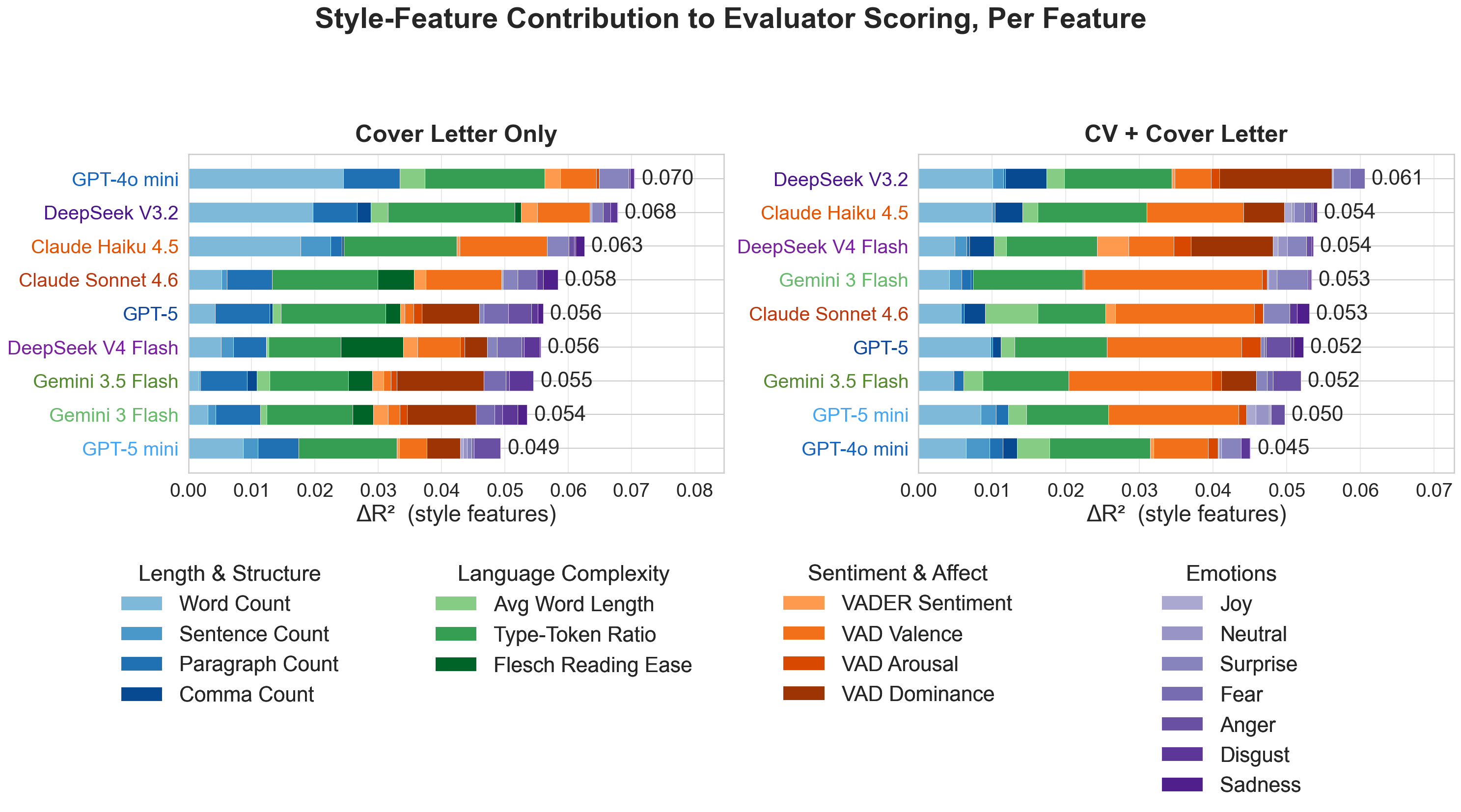}
  \caption{$\Delta R^2$ of individual style features on top of the
    same Job~+~Substance~+~Job-Ad~Fit~+~CV~Consistency baseline as
    Figure~\ref{fig:substance_style}; category-level totals here are
    identical to that figure's style-category bars, this view simply
    decomposes them feature by feature.}
  \label{fig:style_drilldown}
\end{figure}

\paragraph{Style-driven variance, per feature.}
Figure~\ref{fig:style_drilldown} decomposes each evaluator's
style-driven variance (the non-Substance, non-semantic remainder from
Figure~\ref{fig:substance_style}) down to the individual feature
level; its category-level totals are identical to
Figure~\ref{fig:substance_style}'s style bars, so this view simply
answers which specific stylistic cue underlies that remainder, per
evaluator.

\section{Writer Prompt}\label{secD}

The exact prompt template sent to every writer model, verbatim:

\begin{verbatim}
Role: You are a professional career coach who writes tailored,
persuasive cover letters.

Task:
Write a {350-400} word cover letter for this job description:
{job_description}

Using the information given in this CV:
{cv_text}

Output: **DO NOT include any additional commentary, signs or
symbols.** Make sure that all the information is correct, double
check if all the information is accurate with the information in
the CV.
If you are unsure about any information, leave it out.
Make sure the cover letter is professional and concise.
Start the cover letter with "Dear Hiring Manager" and end it with
"Sincerely, Alexis".
DO NOT add any contact information.
\end{verbatim}

The target word count is drawn uniformly at random from 350--400 for
each letter; \texttt{\{job\_description\}} and \texttt{\{cv\_text\}}
are substituted with the specific job posting and candidate CV being
written for. ``Alexis'' is used as a fixed, gender-neutral signoff
name throughout, so that evaluators cannot infer applicant gender from the name.

\FloatBarrier
\end{appendices}

\end{document}